\documentclass[sn-mathphys]{sn-jnl}

\jyear{2023}%

\theoremstyle{thmstyleone}%
\theoremstyle{thmstyletwo}%
\newtheorem{remark}{Remark}%

\theoremstyle{thmstylethree}%

\usepackage{hyperref}
\hypersetup{
    colorlinks=true,
    linkcolor=blue,
    filecolor=magenta,      
}

\usepackage{lineno}
\modulolinenumbers[5]

\usepackage{mathtools}
\usepackage{amssymb}
\usepackage{amsfonts}
\usepackage{amsmath}
\usepackage{lipsum}
\usepackage{graphicx}
\usepackage{epstopdf}
\usepackage[export]{adjustbox}
\usepackage{float}
\usepackage{geometry}
\usepackage{lscape}
\usepackage{color,soul}
\usepackage[english]{babel}

\usepackage{enumitem}
\setlist{nosep}
\usepackage{algorithm}
\usepackage{algpseudocode}
\usepackage{cleveref}
\crefname{algocf}{alg.}{algs.}
\Crefname{algocf}{Algorithm}{Algorithms}

\newcommand{\bs}[1]{\boldsymbol{#1}}
\newcommand{\mc}[1]{\mathcal{#1}}
\newcommand{\eps}[0]{\boldsymbol{\varepsilon}}
\newcommand{\sig}[0]{\boldsymbol{\sigma}}
\newcommand{\tr}[0]{\textrm{tr}}

\begin{document}

\title{An efficient phase-field model of shear fractures using deviatoric stress split}


\author[1]{\fnm{Ehsan} \sur{Haghighat}}

\author[2]{\fnm{David} \sur{Santill{\'a}n}}\email{david.santillan@upm.es}


\affil[1]{\orgdiv{Department of Civil and Environmental Engineering}, \orgname{Massachusetts Institute of Technology}, \orgaddress{\street{77 Massachusetts Ave}, \city{Cambridge}, \postcode{02139}, \state{Massachusetts}, \country{USA}}}

\affil[2]{\orgdiv{Departamento de Ingener{\'i}a Civil: Hidr{\'a}ulica, Energ{\'i}a y Medio Ambiente }, \orgname{Universidad Polit{\'e}cnica de Madrid}, \orgaddress{\street{C/Profesor Aranguren 3},  \city{Madrid}, \postcode{28040}, \country{Spain}}}


\abstract{We propose a phase-field model of shear fractures using the deviatoric stress decomposition (DSD). This choice allows us to use general three-dimensional Mohr-Coulomb's (MC) failure function for formulating the relations and evaluating peak and residual stresses. We apply the model to a few benchmark problems of shear fracture and strain localization and report remarkable performance. Our model is able to capture \emph{conjugate} failure modes under biaxial compression test and for the slope stability problem, a challenging task for most models of geomechanics.}

\keywords{Phase-field, Shear fracture, Strain localization, Slope stability}

\maketitle

\section{Introduction}

The shear failure of brittle materials in compression, also known as shear bands or localized strains, are one of the dominant modes of failure in geo-structures. It has recently emerged as an active research topic due to its interest in structural geology and engineering. The growing interest stems from its engineering applications in subsurface energy technologies, including enhanced geothermal energy systems where the hydro-shearing technique is aimed to reactivate and slide the preexisting fracture network to increase the rock mass permeability \cite{rinaldi2015coupled,rinaldi2019joint,andres2019thermo,andres2022hydraulic}, large-scale CO$_2$ sequestration in deep saline aquifers \cite{vilarrasa2015geologic,juanes2012no,white2016assessing}, impoundment and level changes of artificial water reservoirs of hydropower plants \cite{gupta2002review,mcgarr2002case,rinaldi2020combined,pampillon2020geomechanical} and underground natural gas storage facilities \cite{vilarrasa2021unraveling}, where their mechanics are crucial to understanding the stability of faults and hence earthquake mechanisms \cite{cueto2017stick,cueto2018numerical,andres2019delayed,pampillon2023}. Other engineering applications include fault and slope stability assessment \cite{veveakis2007thermoporomechanics,borja2016rock}, or the stability of faults during the groundwater injection and production operations \cite{gonzalez20122011,tiwari2021groundwater}.

The simulation of shear fracturing processes is a challenging task. The finite element method (FEM) has been the dominant numerical method for modeling solids and continua. Classically, two fundamentally different perspectives are proposed to study compressive fractures using FEM:
\begin{itemize}
\item [-] Discrete fracture models (DFM) that are based on the classical theory of Linear Elastic Fracture Mechanics (LEFM) founded by \citet{griffith1921vi,irwin1956onset}. 
\item [-] Smeared fracture models (SFM) that are based on the classical theory of Continuum Damage Mechanics (CDM) proposed initially by \citet{kachanov1958rupture}.
\end{itemize}
Each class includes extensive literature dating back to the 1960s that is out of the scope of this text to cover comprehensively. Therefore we only point the interested reader to a few primary studies of each class. 

Within the DFM realm, common approaches include node duplication on fracture interface \cite{chan1970finite,rybicki1977finite,bavzant1979blunt}, strong discontinuity approaches \cite{pietruszczak1981finite,simo1987strain,belytschko1988finite,simo1993analysis,simo1994new,oliver2000discrete,regueiro2001plane,wells2001three,foster2007embedded,liu2008contact,dias2009discrete,haghighat2015modeling}, and Extended Finite Element Methods (XFEM) \cite{moes1999finite,dolbow2001extended,moes2002extended,areias2005analysis,song2006method,liu2008contact,borja2008assumed,sanborn2011frictional,mikaeili2018xfem,hirmand2015augmented}. 
These methods require using geometrical algorithms to trace the fracture propagation, which has been found very challenging for generic three-dimensional setups. Such methods are efficient for modeling single fractures. However, they become quickly impractical when dealing with complex fracture nucleation and propagation patterns. 

As per the SFMs, we can point to continuum damage models (CDM) \cite{kachanov1958rupture,bavzant1979blunt,kachanov1986introduction,bavzant1988nonlocal,leroy1989finite,ovzbolt1996numerical,bavzant2002nonlocal}, peridynamic models \cite{silling2000reformulation,kilic2009structural,silling2010peridynamic,agwai2011predicting,madenci2014peridynamic,ren2016new,madenci2016peridynamic,kamensky2019peridynamic,song2019peridynamics,zhang2022peridynamic}, and phase-field models (PFM) \cite{francfort1998revisiting,bourdin2000numerical,bourdin2008variational,miehe2010thermodynamically,miehe2010phase,kuhn2010continuum}, which we discuss next in more details. 
While early models showed significant mesh dependencies, these models have been used to simulate very complex fracture patterns under realistic conditions. Among this class, phase-field models have been most attractive in recent years due to their thermodynamically sound foundations and their ability to model complex fracture patterns.

Phase-field models have been extensively used for modeling brittle, cohesive, and ductile Mode-I fracture patterns, in elastic or poroelastic materials and homogeneous or heterogeneous domains \cite{francfort1998revisiting,bourdin2000numerical,bourdin2008variational,miehe2010thermodynamically,miehe2010phase,kuhn2010continuum,borden2012phase,verhoosel2013phase,borden2014higher,ambati2015phase,santillan2017phase,santillan2017b,santillan2018phase,santillan2017fluid,Aldakheel2021,Seles2021} \citep[see][for a detailed review]{wu2020phase}. 
Although \citeauthor{lancioni2009variational} proposed a simple extension for shear fractures, the applicability of phase-field for modeling shear failure remained virtually untouched until very recently \cite{bryant2018mixed,zhou2019phase,fei2020phase}. In a detailed study, \citeauthor{fei2020phase} presented a phase-field formulation of frictional fracture based on \citeauthor{palmer1973growth} theory and using a similar stress decomposition approach to the one proposed by \citeauthor{hu2020phase} for tensile cracks. The authors validated their model on a set of classical problems as well as various experimental setups \cite{fei2021double}.

In the present study, we propose a phase field model of shear failure that adapts the cohesive model of shear fractures proposed by \citeauthor{fei2020phase} for deviatoric stress decomposition (DSD) instead of the proposed contact stress decomposition (CSD). Hence, we arrive at an alternative descriptor for the shear fracture orientation (i.e., the $\boldsymbol{\alpha}$ tensor) which is solely based on the deviatoric strain. We adapt the crack driving force to be consistent with the DSD decomposition. The resulting formulation simplifies the damage criterion since it results in damaging the shear modulus. Lastly, the proposed model allows us to use the general forms of the failure functions from the classical plasticity theory and therefore is not limited to Mohr-Coulomb failure model.

In what follows, we first briefly describe the original framework based on CSD. We then discuss our generalization proposal. Lastly, we use both frameworks to model a set of benchmark problems.

\section{Phase-field method}
In this section, we first describe the general phase-field framework for modeling crack propagation in materials. We then summarize the most recent CSD shear model \cite{fei2020phase}. Finally, we discuss our proposed adjustment for better stability.

\begin{figure}[t]
    \centering
    \includegraphics[width=0.5\textwidth]{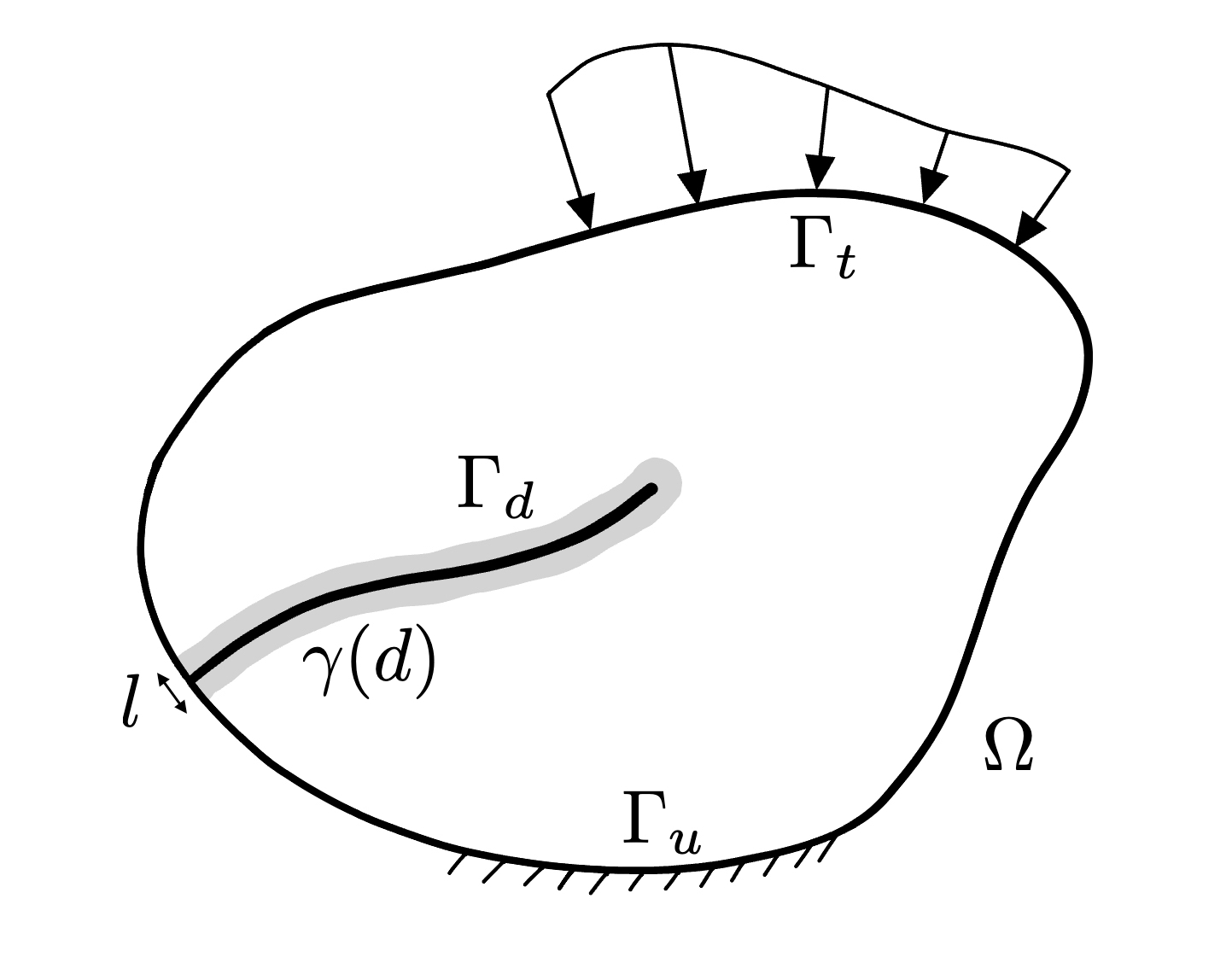}
    \caption{Domain $\Omega$ with boundary $\Gamma$, Dirichlet boundary $\Gamma_u$, and Neumann boundary $\Gamma_t$. The discontinuity surface is represented by $\Gamma_d$ with its phase-field diffused representation as $\gamma(d)$. }
    \label{fig:fig1}
\end{figure}

\subsection{Phase-field governing equations}

Consider the continua $\Omega \in \mathbb{R}^D$ in $D$-dimensional space, depicted in \Cref{fig:fig1}, with its boundary represented as $\Gamma$. The boundary $\Gamma$ is subjected to Neumann boundary conditions on $\Gamma_t$ and Dirichlet boundary conditions on $\Gamma_u$, where $\Gamma_u \cup \Gamma_t = \Gamma$ and $\Gamma_u \cap \Gamma_t = \varnothing$. The set of discontinuities in the domain is represented by a discrete surface $\Gamma_d$. 

According to the phase-field formulation, the fracture's discrete surface $\Gamma_d$ is approximated implicitly as a continuous function over a width $l$ using the Allen-Cahn fracture surface density function $\gamma(d)$ as
\begin{equation}
    \gamma(d) = \frac{1}{c_0 l} \left( w(d) + l^2 \| \nabla d \|^2 \right), \quad \text{with} \quad c_0 = 4\int_0^1 \sqrt{w(l)} dl
\end{equation}
where $d$ is the phase-field variable, with $d=0$ presenting the intact part of the domain while $d=1$ expressing a point on $\Gamma_d$. 
$w(d)$ is the transition function, also known as the dissipation function, defined for cohesive cracks as $w(d)=d$ \cite{kuhn2015degradation,geelen2019phase}, hence $c_0=\frac{8}{3}$. Accordingly, a surface integral $\int ds$ is approximated using a volume integral as $\int ds \approx \int \gamma(d)~dv$.

Given the displacement field $\bs{u}$, the small-deformation strain measure $\eps=(\nabla\bs{u} + \nabla\bs{u}^T)/2$, and the crack surface density function $\gamma(d)$, the total energy of a fractured continua, occupying the domain $\Omega$ and bounded by the boundary $\Gamma$, shown in \Cref{fig:fig1}, is expressed as 
\begin{align}
    \Psi = -\Psi^{external} + \Psi^{internal} + \Psi^{fracture},
\end{align}
where, $\Psi^{external}$ is the work done by the external traction stress $\bs{\tau}$ and body force $\bs{b}$, and expressed as 
\begin{align}
\Psi^{external} &= \int_{\Gamma_t} \bs{u}\cdot\bs{\tau}~ds + \int_{\Omega} \bs{u}\cdot\bs{b}~dv. 
\end{align}
The fracture energy, i.e., $\Psi^{fracture}$, is the energy dissipated from the system to create a fracture surface $\Gamma_d$. Given the energy release rate $\mc{G}_c$ (per unit fracture length), $\Psi^{fracture}$ is expressed as  
\begin{align}
\Psi^{fracture} &= \int_{\Gamma_d} \mc{G}_c ~ds \approx \int_{\Omega} \mc{G}_c \gamma(d)~dv.
\end{align}

The stored internal energy of the system $\Psi^{internal}$ consists of the elastic stored energy in the intact part of the domain and stored energy in the damaged part of the domain, expressed as 
\begin{align}
\Psi^{internal} &= \int_{\Omega} \psi(\eps, d)~dv.
\end{align}
The internal energy density function $\psi(\eps, d)$ is defined as $\psi(\eps,d) = \tfrac{1}{2} \sig : \eps$, which consists of both inactive and damaged counterparts. For the intact part of continuum, i.e., where $d=0$, the Cauchy stress tensor ${\sig}(\eps, d=0)$ is expressed using Hook's law as
\begin{align}
{\sig}(\eps, d=0) = (\kappa-\frac{2}{3}\mu) \varepsilon_v \bs{1} + 2\mu\eps,
\end{align}
where, $\kappa$ and $\mu$ are bulk and shear moduli of the intact material, respectively, and $\varepsilon_v$ is the volumetric strain, expressed as $\varepsilon_v = \tr(\eps)$. For the parts of the domain where $d > 0$, the Cauchy stress tensor is  decomposed into inactive part $\sig^I$ and active part $\sig^A$ as 
\begin{align}
\begin{split}
\sig(\eps, d) &= \sig^I(\eps) + \sig^A(\eps,d) \\ &= \sig^I(\eps) + g(d) \hat{\sig}(\eps) + (1-g(d))\tilde{\sig}(\eps).
\end{split}
\end{align}
The active part of the stress tensor undergoes the damage process, and $g(d)$ is a degradation function that expresses the stress transition from bulk ($\hat{\sig}$) to fracture ($\tilde{\sig}$). We will discuss these in more details in the next sections. 

Therefore, there are two solution variables associated with the phase-field formulation, the standard displacement field $\bs{u}$ and the additional phase-field variable $d$. Taking the variation of $\Psi$ with respect to $\bs{u}$ and $d$, and following the standard \emph{weak to strong} form steps of the FEM \cite{hughes_fem_book_2012,belytschko_fem_book_2014} and phase-field \cite{santillan2017phase,hu2020phase,fei2020phase}, we can arrive at the following governing relations: 

\begin{align}
& \nabla\cdot \sig(\eps,d) + \bs{b} = 0, \label{pde_u}\\
& \frac{3\mc{G}_c}{8l}\left(2l^2\nabla^2 d - 1\right) - g'(d) \mc{H}(\eps) = 0. \label{pde_d}
\end{align}
The irreversibility of the fracture process is guaranteed with the local history field of maximum stored shear energy $\mc{H}^+(\eps)$ that allows us to solve the constrained minimization of \cref{pde_d} in a straightforward way \citep{miehe2010phase} and avoids unphysical self-healing. $\mc{H}^+(\eps)$ is defined as follows:
\begin{equation}
    \mc{H}^+(\eps) = \max_{s\in[0,t]}\left( \mc{H}(\eps) \right),
\end{equation}
where $t$ is time. \Cref{pde_d} is then rewritten as follows:
\begin{equation}
    \frac{3\mc{G}_c}{8l}\left(2l^2\nabla^2 d - 1\right) - g'(d) \mc{H}^+(\eps) = 0.
\end{equation}
Since $\dot{\mc{H}}\geq0$, non-negative $\dot{d}$ is guaranteed and, consequently, the irreversibility of the fracture growth. We define $\mc{H}(\eps)$ after describing the stress decomposition approach. 

In this work, we use the Lorenz degradation function $g(d)$ defined as  \cite{lorentz2011convergence,lorentz2017nonlocal}:
\begin{equation}
    g(d, p) = \frac{(1-d)^2}{(1-d)^2 + \frac{M}{\psi_c}d(1+p d)}
\end{equation}
where, $M={G_c}/({c_0 l})$ and $\psi_c$ is the critical crack driving force at the material's peak strength, evaluated as $\psi_c = -M w'(0)/{g'(0)}$. The damage begins to accumulate as soon as elastic stored energy exceeds this critical threshold. Here, we take $p=1$.



\subsection{Stress decomposition. Introduction}

The split of the strain energy density into crack driving and intact components defines the damage mode and fracture pattern. Up to date, two fundamental approaches are available. 
The approaches of the first class do not take into account the local fracture orientation, whereas the second approaches take into consideration the local crack orientation.

The first group of models includes the isotropic model, the volumetric and deviatoric decomposition model, the spectral decomposition model, or the anisotropic models. The isotropic model proposed by \citeauthor{bourdin2000numerical} where the entire strain energy density is degraded. The volumetric and deviatoric decomposition model proposed by \citeauthor{amor2009regularized} splits the strain tensor into its volumetric and deviatoric components. This approach avoids crack inter-penetration in composites and masonry structures. The fracture is then assumed to be driven by volumetric expansion and deviatoric strains. The spectral decomposition model proposed by \citeauthor{miehe2010phase} splits the strain tensor into its principal components and only tensile components drive the fracture propagation. The anisotropic models are based on the spectral decomposition of the strain tensor using other projections, such as the eigenvalue and eigenvector of the effective stress tensor \cite{wu2020variationally}.

The second group of approaches take into consideration the local crack orientation. The directional model proposed by~\citeauthor{steinke2019phase} splits the stress tensor into the crack driving and persistent components using the fracture orientation. For each point, a fracture coordinate system is defined and the fracture orientation is obtained from the maximum principal stress direction. ~\citeauthor{strobl2016constitutive} and~\citeauthor{strobl2015novel} computed the fracture orientation from the phase-field gradients. Following this way to compute the fracture direction,~\citeauthor{liu2021micromechanics} developed a phase field model based on micromechanical modeling,~\emph{i.e.}, the macroscopic fracture is modeled as a collection of microscopic fractures.

In the following subsections, we describe the contact stress decomposition (CSD), used satisfactorily to simulate shear fractures under confining pressures, and lastly we present our proposal based on the deviatoric stress decomposition (DSD). 
Both models do not take into account the local fracture orientation.

\subsubsection{Contact stress decomposition (CSD)}

Since a compressive fracture behaves like a contact problem, \citeauthor{fei2020phase} proposed a stress decomposition approach that is closely related to the contact formulation, which we refer here as CSD. It starts by considering a corotational coordinate system on the fracture surface with $\bs{m}$ and $\bs{n}$ as tangential and normal vectors to the crack surface, and $\bs{m}$ along the direction of sliding. Additionally, let us define $\bs{\alpha} = (\bs{m}\bs{n} + \bs{n}\bs{m})/2$. 

According to this approach and under the assumption that the fracture remains closed, i.e., no tensile fracture, the only stress component that should undergo damage is the shear stress, and other stress components remain inactive. The bulk shear stress can be expressed as  
\begin{align}
\hat{\tau} &= \hat{\sig}:\bs{\alpha} = \mu\varepsilon_{\gamma},
\end{align}
where, $\varepsilon_{\gamma} = 2~\eps:\bs{\alpha} = 2~\bs{m}\cdot\eps\cdot\bs{n}$. Consider the contact shear stress as $\tilde{\tau}$. Then, the inactive stress tensor is expressed as 
\begin{align}
\sig^I = \sig(\eps, d=0) - \mu\varepsilon_{\gamma}~\bs{\alpha},
\end{align}
and the active stress tensor as 
\begin{align}
\sig^A = \tau^A \bs{\alpha}, \quad \text{where} \quad \tau^A = \mu\varepsilon_{\gamma}~g(d)+ \tilde{\tau}~(1-g(d)).
\end{align}
Here, $\tilde{\tau}$ is the residual contact stress while the fracture is fully developed, i.e., d=1. 


\begin{remark}
Given the Mohr-Coulomb's failure function as, 
\begin{equation}
\mc{F} = \lvert \tau \rvert - \sigma_n\tan\phi - c = 0
\end{equation}
with $\sigma_n=\bs{n}\cdot\sig\cdot\bs{n}$ as to normal stress on the fracture surface, and $c$ and $\phi$ as cohesion and friction angle of the intact material, the peak and residual shear stresses are expressed as 
\begin{align}
\tilde{\tau}_p = c + \sigma_n \tan(\phi), \quad
\tilde{\tau}_r = c_r + \sigma_n \tan(\phi_r),
\end{align}
where $c_r$ and $\phi_r$ are residual friction and cohesion at the fully developed failure state.
\end{remark}

\begin{remark}
Based on the Mohr-Coulomb failure criterion, the critical plane for the failure is evaluated at two conjugate angles $\theta = \pm(45^\circ - \phi_r/2)$ \cite{pietruszczak1981finite} with respect to the direction of the maximum principal stress. However, the authors only consider $\theta = +(45^\circ - \phi_r/2)$ \citep[see][eq.56]{fei2020phase}. This restriction is required otherwise $\bs{m},\bs{n}$ is not uniquely defined. 
\end{remark}



\subsubsection{Our proposal: deviatoric stress decomposition (DSD)}

The total strain tensor can be decomposed into volumetric and deviatoric parts, as $\eps = \varepsilon_v\bs{1} + \bs{e}$. We can also express the Cauchy tensor in terms of the mean confining stress $p$ and the deviatoric stress tensor $\bs{s}$ as $\sig = -p\bs{1} + \bs{s}$. Therefore, we can  re-write Hook's law for the intact part as \cite{lancioni2009variational,zhang2022assessment} 
\begin{align}
\sig(\eps, d=0) 
= -p\bs{1} + \bs{s} = \kappa ~ \varepsilon_v \bs{1} + 2\mu\bs{e}.
\end{align}
Given the equivalent deviatoric (Mises) stress ${q} = (\frac{3}{2} {\bs{s}}:{\bs{s}})^{1/2}$ and the equivalent deviatoric strain $\varepsilon_q = (\frac{2}{3}\bs{e}:\bs{e})^{1/2}$ and with some algebra, we can write that 
\begin{align}
{q} = 3\mu \varepsilon_q.
\end{align}
Let us now define the \emph{Unit Deviator Tensor} $\bs{\alpha}_q$ as
\begin{align}
    \bs{\alpha}_q = \sqrt{\frac{2}{3}} \frac{\bs{e}}{\varepsilon_q},\quad \text{where} \quad \|\bs{\alpha}_q\| = \sqrt{\bs{\alpha}_q:\bs{\alpha}_q} = 1,
\end{align}
Hook's law can therefore be expressed as 
\begin{align}
{\bs{\sigma}} = -{p} \bs{1} + \sqrt{\frac{2}{3}} {q} \bs{\alpha}_q, \quad \text{where} \quad {p} = \kappa\varepsilon_v, ~~{q} = 3\mu\varepsilon_q.
\end{align}

Equivalent to the CSD, we can describe the compressive failure in a material as damage in the deviatoric stress component. Therefore, the compressive pressure becomes the inactive part of the stress tensor, i.e.,
\begin{align}
{\sig}^I = -{p} \bs{1} = \kappa\varepsilon_v\bs{1},
\end{align}
and active stress is described as 
\begin{align}
\sig^A = q(\eps, d) \bs{\alpha}_q, \quad \text{where} \quad q(\eps, d) = g(d) \hat{q} + (1-g(d)) \tilde{q}_r,
\end{align}
where the bulk deviatoric stress is $\hat{q} = 3\mu \varepsilon_q$.

\begin{remark}
This deviatoric stress decomposition allows us to leverage the general form of virtually any failure surface that are described in the classical plasticity theory, including the Mohr-Coulomb failure function. Given the friction angle $\phi$ and cohesion coefficient $c$, the general form of the Mohr-Coulomb's failure criterion is expressed as
\begin{align}
\mc{F} = \mc{R}_{MC} {q} - {p}\tan\phi - c = 0.
\end{align}
Here, $\mc{R}_{MC}$ defines the shape of the Mohr-Coulomb's failure surface and is expressed as 
\begin{align}
\mc{R}_{MC} = \frac{1}{\sqrt{3}\cos\phi} \sin(\Theta + \frac{\pi}{3}) + \frac{1}{3}\cos(\Theta + \frac{\pi}{3})\tan\phi,
\end{align}
where $\Theta$ is the Lod\`e angle, evaluated as $\cos(3\Theta) = \left({r}/{q}\right)^3$. The invariant ${r}$ is the third invariant of the deviatoric stress tensor, and is defined as ${r} = (\frac{9}{2}tr({\bs{s}}^3))^{1/3}$. 
Based on this criterion, we can find the peak and residual Mises stresses as 
\begin{align}
    \tilde{q}_{p} = \frac{{p}\tan\phi + c}{\mc{R}_{MC}}, \quad \tilde{q}_{r} = \frac{{p}\tan\phi_r + c_r}{\mc{R}_{MC}},
\end{align}
with $\phi_r$ and $c_r$ as the residual values for friction angle and cohesion at the fully damaged state. 
\end{remark}

\begin{remark}
We can easily replace the non-smooth Mohr-Coulomb surface $\mc{R}_{MC}$ with some alternatives \cite{pietruszczak_plasticity_book_2010,borja_plasticity_book_2013}. In fact, we can potentially pick any alternative failure function available for different materials. 
\end{remark}


\subsection{Crack driving force}

Given $\tau = \mu \varepsilon_{\gamma}$ and $\tau_p = p \tan\phi + c = \mu \varepsilon_{\gamma}^p$, the crack driving force relations for CSD is derived as \cite{fei2020phase}
\begin{align}
\mc{H} = \mc{H}_t + \mc{H}_{slip}
\end{align}
where
\begin{align}
&\mc{H}_t = \frac{\tilde{\tau}_p - \tilde{\tau}_r}{2\mu}, \\
&\mc{H}_{slip} = \frac{1}{2\mu} \left[ (\hat{\tau} - \tilde{\tau}_r)^2 - (\tilde{\tau}_p - \tilde{\tau}_r)^2 \right]
\end{align}
and they showed that this model is consistent with \citeauthor{palmer1973growth} model. 
Now, for the deviatoric stress decomposition discussed above, we can revise the crack driving force, given $\hat{q} = 3\mu\varepsilon_q$ and $\tilde{q}_p = (p \tan\phi + c)/\mc{R}_{MC} = 3\mu\varepsilon_q^p$, as
\begin{align}
&\mc{H}_t = \frac{(\tilde{q}_p - \tilde{q}_r)^2}{6\mu}, \label{eqs:28}\\
&\mc{H}_{slip} = \frac{1}{6\mu} \left[ (\hat{q} - \tilde{q}_r)^2 - (\tilde{q}_p - \tilde{q}_r)^2 \right] \label{eqs:29}.
\end{align}
More details on the derivation of $\mc{H}_t$ and $\mc{H}_{slip}$ for CSD approach are provided in \Cref{appen1}.

\subsection{Boundary conditions}
To have a complete mathematical description of the problem, we lastly need to describe the boundary conditions. Considering \Cref{fig:fig1}, the boundary conditions are described as 
\begin{align}
    \bs{u} = \bar{\bs{u}}, \quad  &\text{on} \quad \Gamma_u, \\
    \bs{\tau} = \sig\cdot\bs{n} = \bar{\bs{\tau}}, \quad &\text{on} \quad \Gamma_t, \\
    \nabla d\cdot\bs{n} = 0, \quad &\text{on} \quad \Gamma,
\end{align}
where $\bar{\bs{u}}$ and $\bar{\bs{\tau}}$ are prescribed displacement and traction forces, respectively. 

The steps used to solve the problem are detailed in Algorithm 1.

\begin{algorithm}\label{pseudocode}
\caption{Pseudo-code for DSS phase-field model of shear fractures}\label{alg:cap}
\begin{algorithmic}[1]
\State $\bs{u}^0, d^0 \gets 0 $
\State $\bs{\sigma}^0 \gets $ Initial stress using an static step
\State $t = 0$
\For{time-steps}
\State $t \gets t + \Delta t$
\State $\bar{\bs{u}}^{t + \Delta t}, \bar{\bs{\tau}}^{t + \Delta t} \gets $ Update displacement/traction BCs at $t + \Delta t$ 
\State $\bs{u}_0, \bs{\varepsilon}_0, \bs{\sigma}_0, \bs{d}_0 \gets \bs{u}_0^t, \bs{\varepsilon}_0^t, \bs{\sigma}_0^t, \bs{d}_0^t$
\While{$ \text{err} > \text{TOL}$}
\State $\delta \bs{u} \gets$ {Solve \cref{pde_u} for displacement increment}\
\State $\bs{u} \gets \bs{u} + \delta \bs{u}$
\State $\bs{\varepsilon}, \varepsilon_v, \varepsilon_q \gets \bs{\varepsilon} + \delta \bs{\varepsilon}$
\State $\bs{\alpha}_q \gets \sqrt{2/3} ~\bs{e}/\varepsilon$
\State $\hat{q} \gets 3\mu\varepsilon_q$
\State $\tilde{q} \gets (p \tan \phi_r + c_r) / \mc{R}_{MC}$
\State $\bs{\sigma} \gets -\kappa \varepsilon_v \bs{1} + \left[g(d)\hat{q} + (1-g(d)) \tilde{q}_r \right]~\sqrt{2/3}\bs{\alpha}_q$
\State $\mathcal{H}_t \gets (\tilde{q}_p - \tilde{q}_r)^2 / 6\mu$
\State $\mathcal{H}_{slip} \gets \max(\mathcal{H}_{slip}^{t}, ~\left[(\hat{q} - \tilde{q}_r)^2 - (\tilde{q}_p - \tilde{q}_r)^2 \right]/6\mu)$
\State $\mc{H}^+ \gets \mc{H}_t + \mc{H}_{slip}$
\State $\delta d \gets$ {Solve \cref{pde_d} for \textit{d}}\
\EndWhile
\EndFor
\end{algorithmic}
\end{algorithm}

\section{Applications to compressive strain localization}

Here, we consider three reference problems of shear fractures, including direct shear test, biaxial compression test, and slope failure analysis. We show that our model can effectively capture multiple modes of failure concurrently.

\subsection{Direct shear test}

\begin{figure*}
    \centering
    \noindent\includegraphics[trim = 23mm 166mm 15mm 60mm, clip, width=5in]{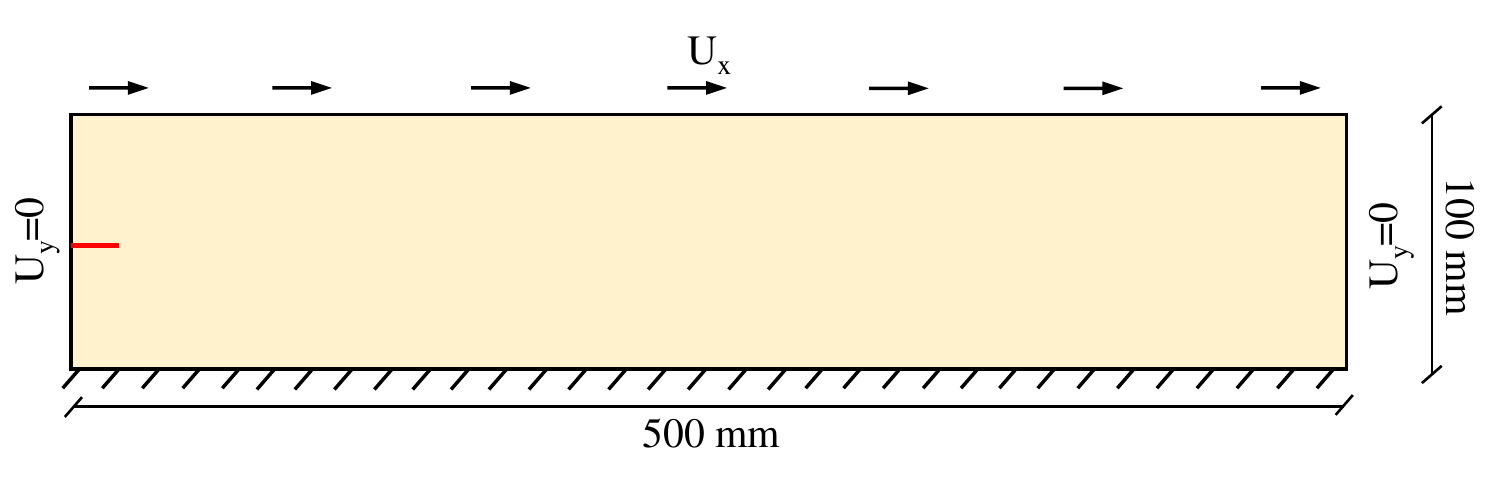} 
	\caption{Direct shear test setup. The domain is 500 mm long, 100 mm tall, and an initial 10-mm horizontal fracture is carved in the middle of the left boundary --red fracture--. The boundary conditions are: the bottom boundary is fixed, the top boundary is displaced horizontally, and the two lateral boundaries are fixed vertically.}\label{Fig5}
\end{figure*}

Our first example is the direct shear test. We simulate the propagation of a fracture in a long shear apparatus and we compare our results with analytical solutions and Fei and Choo's numerical simulations~{\cite{fei2020phase}}. 
The setup of the experiment is plotted in \Cref{Fig5}. The domain is 500 mm long, 100 mm tall, and an initial 10-mm horizontal fracture is carved in the middle of the left boundary. The boundary conditions are: the bottom boundary is fixed, the top boundary is displaced horizontally, and the two lateral boundaries are fixed vertically. We neglect gravity. 

The material properties are: shear modulus $G=10$ MPa, Poisson's ratio $\nu=0.3$, cohesion strength $c=40$ kPa, peak and residual friction angle $\phi=\phi_r=15^{\circ}$, shear fracture energy $\mathcal{G}_{c}=30$ J$/$m$^2$, and fracture's length-scale $l=2$ mm. As in the previous works of \citeauthor{palmer1973growth} and \citeauthor{fei2020phase}, we impose the fracture propagation to be horizontal. Following Fei and Choo's simulations~\cite{fei2020phase}, we initialize vertical compressive normal stress to 149 kPa, which results in $\tau_p=80$ kPa and $\tau_r=40$ kPa. We mesh the domain near the fracture path with a mapped squared mesh of size $l/4=0.5$ mm and the remaining domain with a 1-mm free triangular mesh.

\begin{figure*}
    \centering
	\noindent\includegraphics[trim = 31mm 60mm 31mm 86mm, clip, width=3.0in]{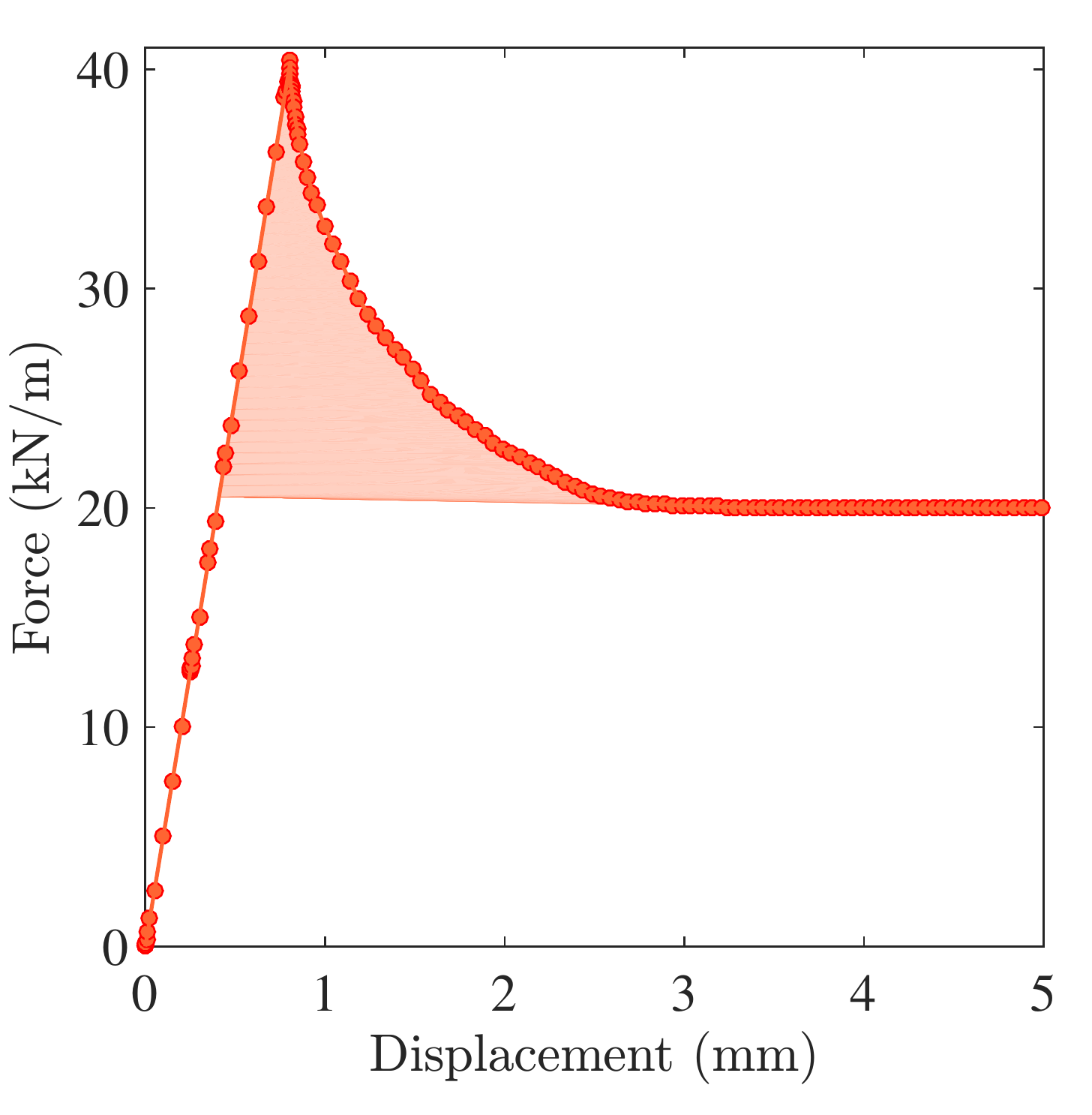} 
	\caption{Horizontal force-displacement curves for the direct shear test. Points are the output of our numerical simulation.}\label{Fig6}
\end{figure*}

The horizontal force-displacement curve is shown in \Cref{Fig6}. The agreement of the peak and residual forces provided by our numerical simulation is very satisfactory.
Theoretically, the peak load, i.e., the peak shear stress times the width of the specimen, is 40 kN, and the output of our simulation is $40.387$ kN. In the same way, the theoretical residual load is 20 kN and the output of our simulation is $19.978$ kN. We estimate the fracture energy from the force-displacement curve, the shaded area in \Cref{Fig6}. The output of our model provides a fracture energy equal to $14.6914$ J, while the theoretical value is 15 J. Therefore, we report a remarkable agreement between our simulations and expected theoretical values.

\begin{figure*}
    \centering
	\makebox[2.3in][l] {\hspace*{0.1in}\textbf{(a)}} \makebox[2.3in][l] {\hspace*{0.1in}\textbf{(b)}}\\
    \noindent\includegraphics[trim = 32mm 65mm 37mm 85mm, clip, width=2.3in]{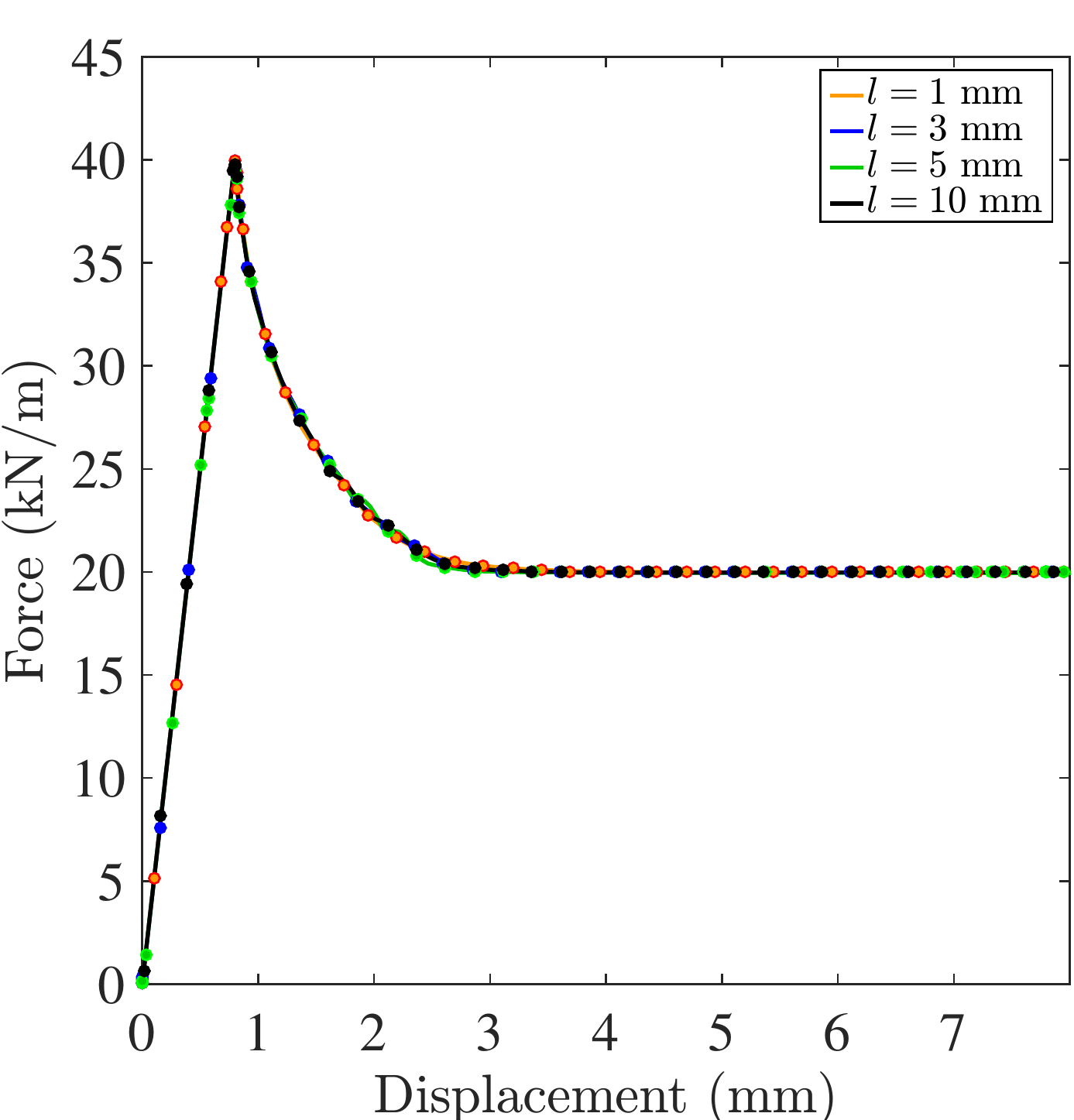} 
	\noindent\includegraphics[trim = 32mm 65mm 37mm 85mm, clip, width=2.3in]{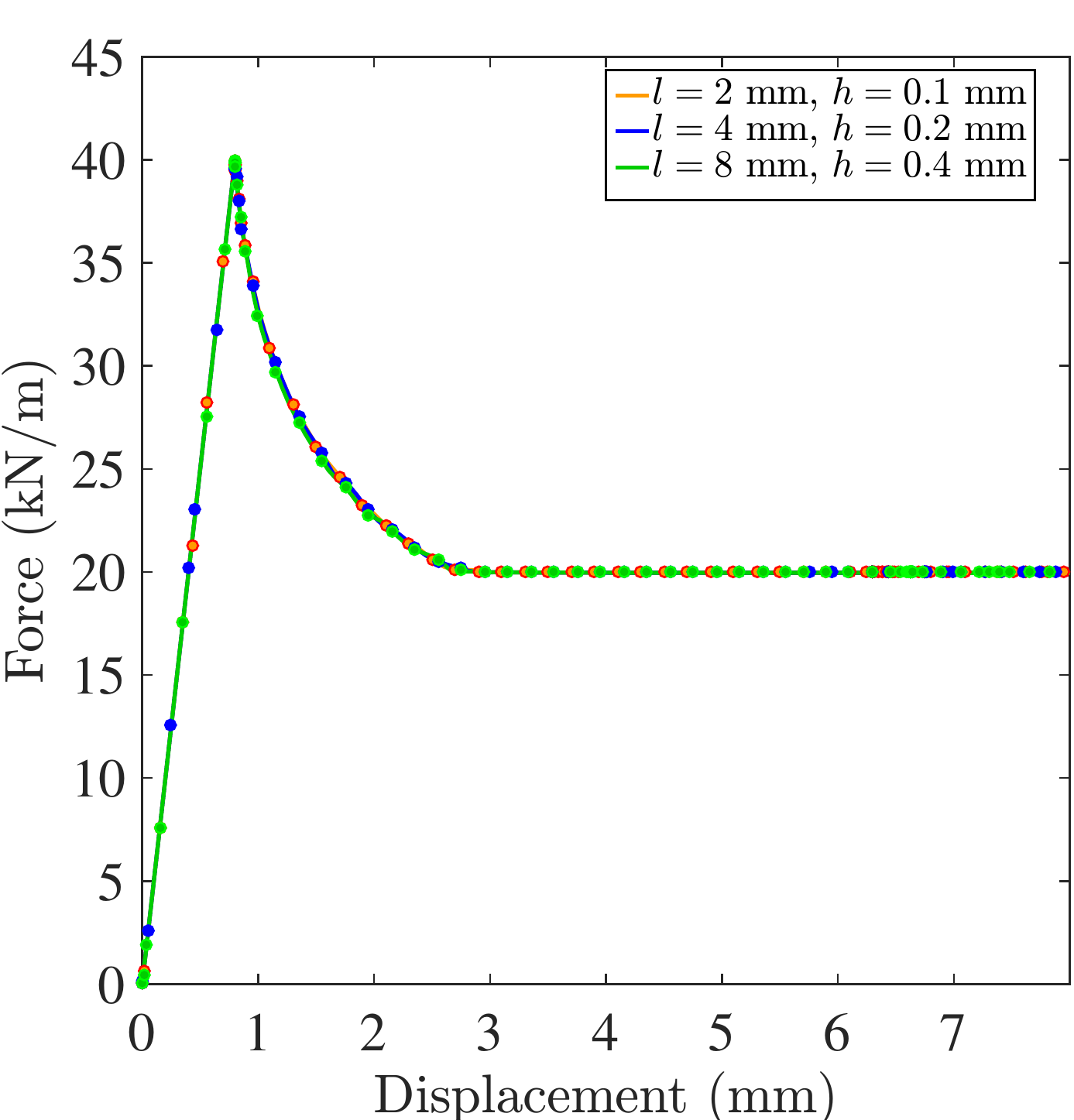} 
	\caption{Force-displacement curves for the direct shear test with several phase-field length parameters, $l$, and mesh sizes, $h$. \textbf{(a)} Here the phase-field length parameter ranges from 1 mm to 10 mm and the mesh size is set to $h=0.2$ mm. \textbf{(b)} Here the ratio phase-field length parameter to mesh size is set to 20.}\label{Fig6_rev}
\end{figure*}

We analyze the sensitivity of our model to the phase-field length parameter, $l$. We run several simulations of the direct shear test problem for several values of $l$, ranging from 1 mm to 10 mm. Results are depicted in \Cref{Fig6_rev}(a). The force-displacement curves for the four values of $l$ confirm that the model is virtually insensitive to the phase-field length parameter. We check the mesh dependency of our model by running three problems of the long-shear apparatus problem. We fix the ratio length scale parameter to mesh size, $l/h$, to 20 and we run three simulations for three $l$- and $h$-values. Results are plot in \Cref{Fig6_rev}(b). The curves confirm that the model is insensible to the mesh size.

\begin{figure*}
    \centering
		\makebox[5in][l] {\hspace*{0.1in}\textbf{(a)}~$U_x=1$ mm }  \\
        \noindent\includegraphics[trim = 34mm 44mm 34mm 44mm, clip, width=5in]{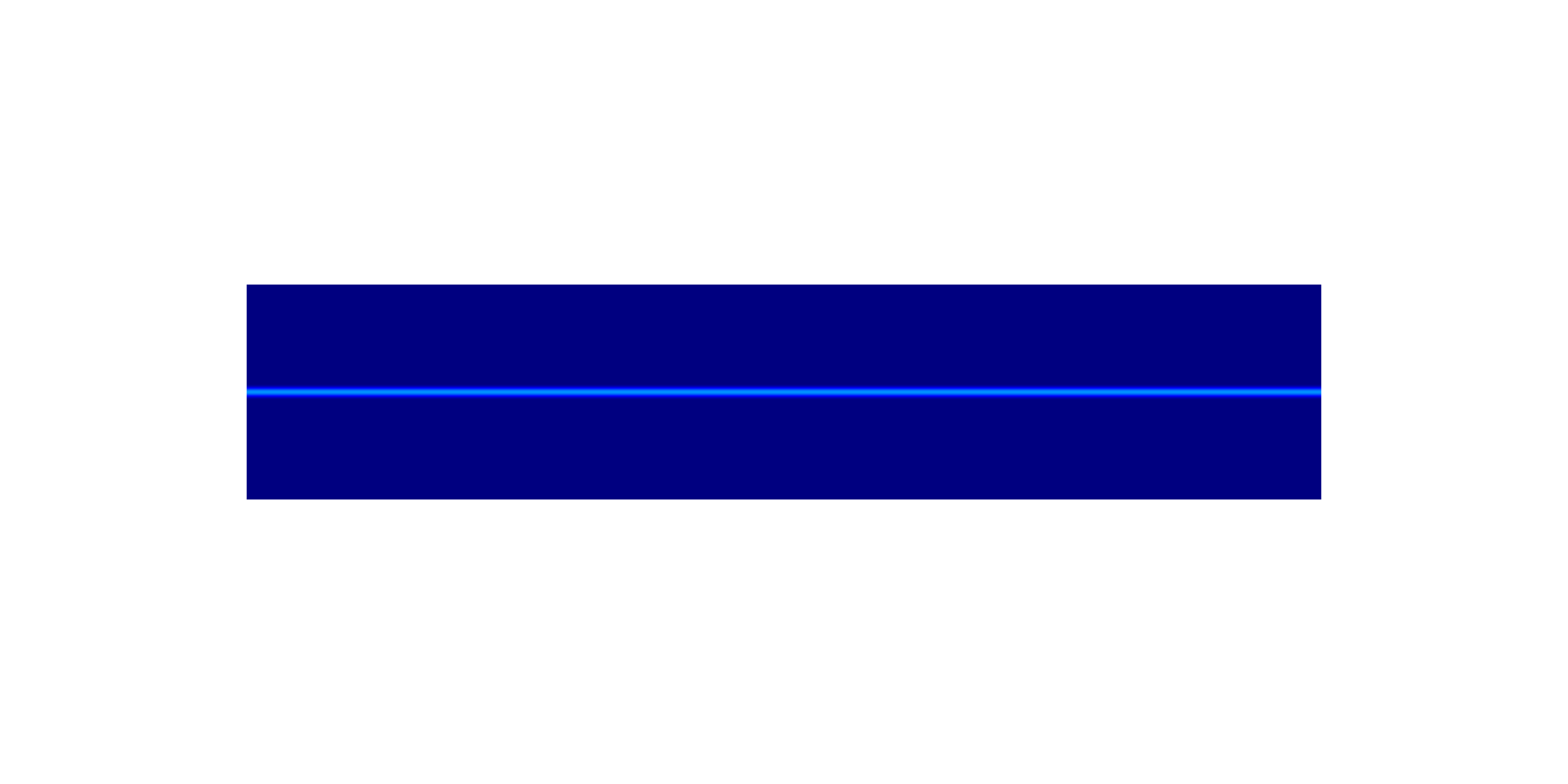} 
		\makebox[5in][l] {\hspace*{0.1in}\textbf{(b)}~$U_x=2$ mm} \\
		\noindent\includegraphics[trim = 34mm 44mm 34mm 44mm, clip, width=5in]{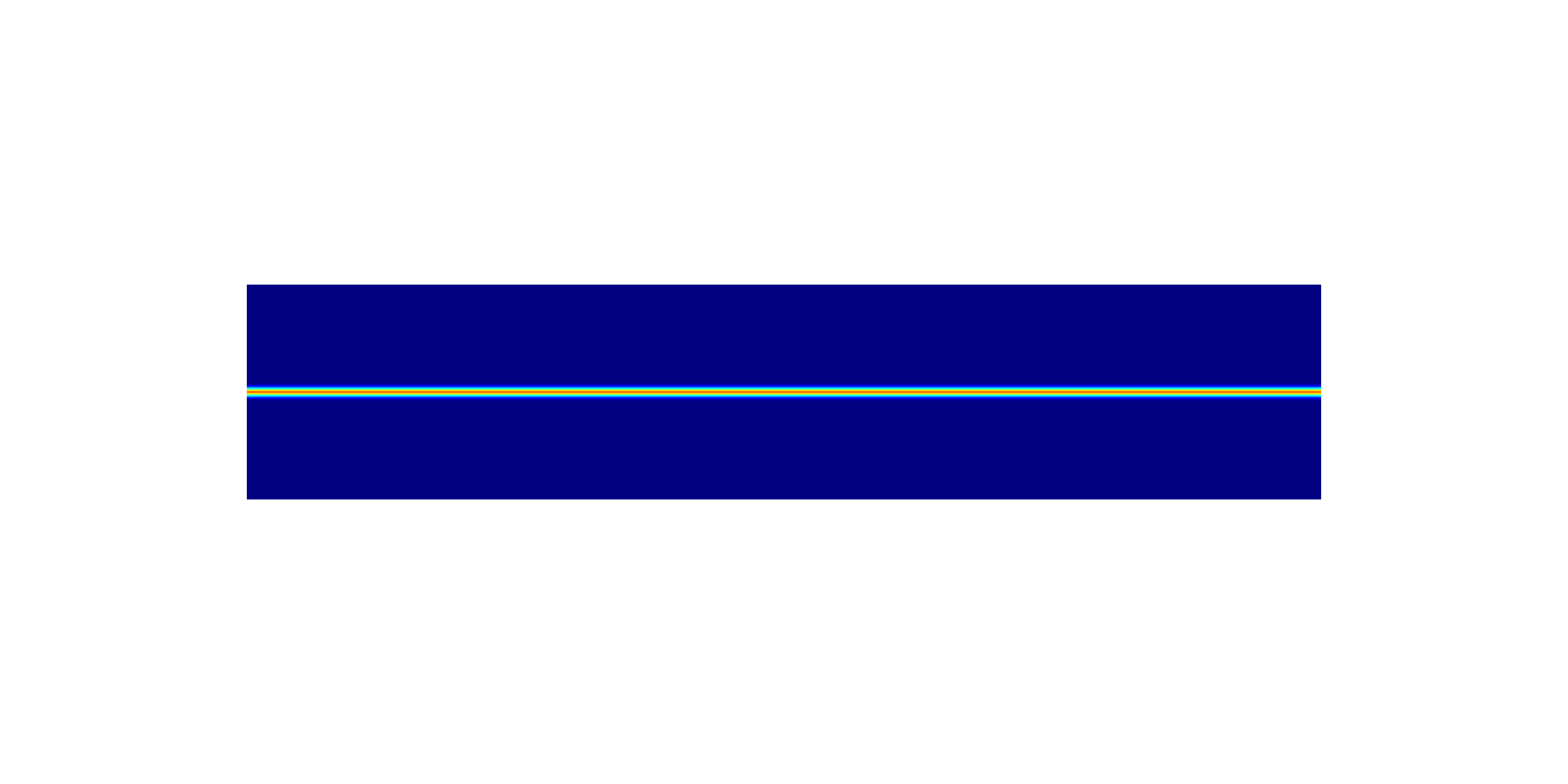} 
		\makebox[5in][l] {\hspace*{0.1in}\textbf{(c)}~$U_x=3$ mm}\\
		\noindent\includegraphics[trim = 34mm 44mm 34mm 44mm, clip, width=5in]{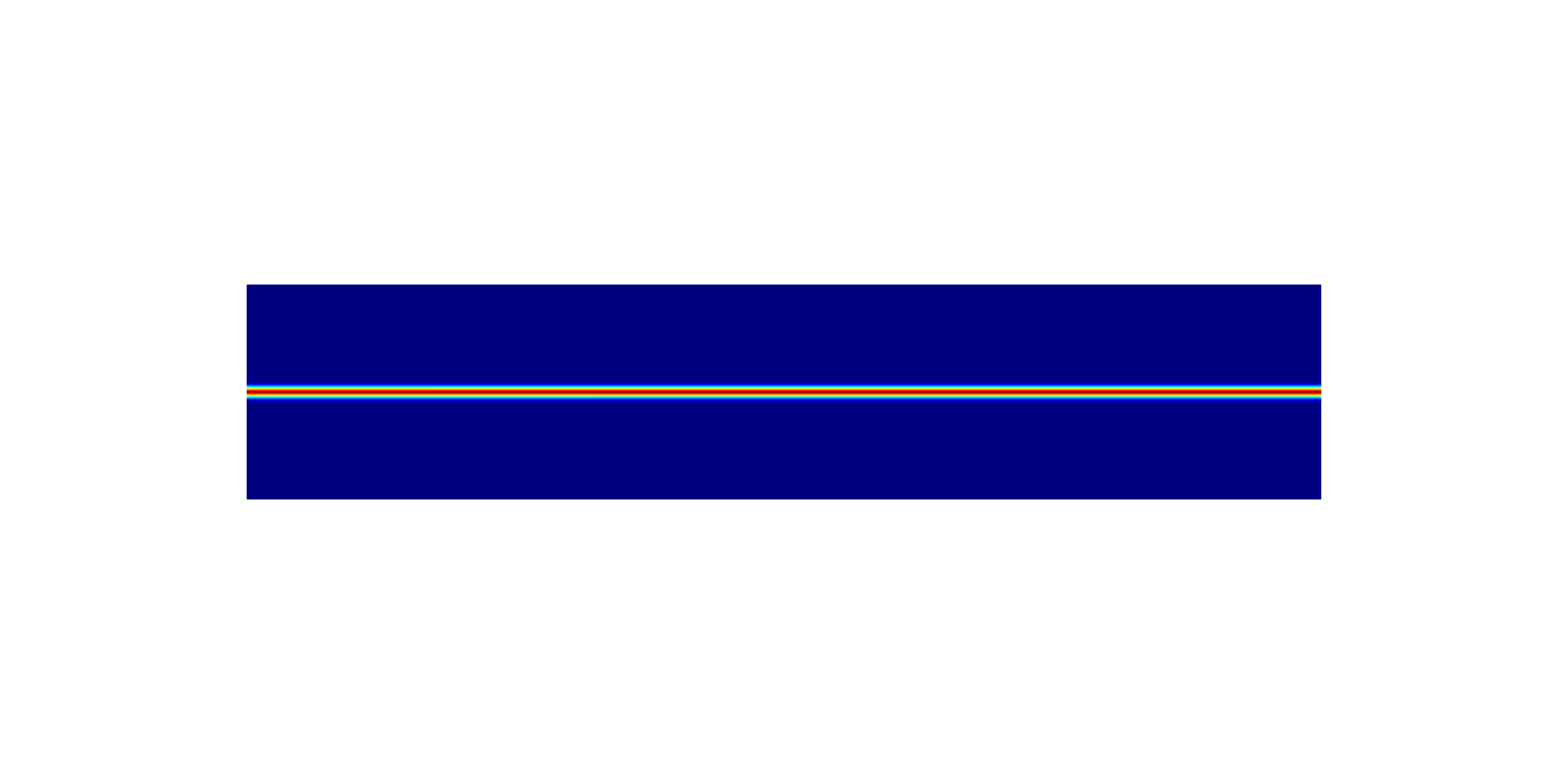} 
        \caption{The evolution of the phase-field variable at three time steps for the direct shear test. The imposed horizontal displacements, $U_x$, are: \textbf{(a)} $1$ mm, \textbf{(b)} $2$ mm, and \textbf{(c)} $3$ mm.}\label{Fig7}
\end{figure*}

We plot the phase-field distribution at three time steps in \Cref{Fig7}. The peak load is given for $U_x=0.8083$ mm, after this value is reached the phase-field has already emerged and propagate along the whole fracture, \Cref{Fig7} (a). Afterward, the phase-field value intensifies during the softening stage, \Cref{Fig7} (b), up to the time the fracture is completely developed, \Cref{Fig7} (c). At this time, the domain is split into two parts. The upper part slips over the bottom one, and the shear stress between both parts is constant and equal to the residual shear stress, $\tau_r=40$ kPa, resulting in a theoretical horizontal force of $20$ kN.   

\subsection{Biaxial compression test}
Our next example is a biaxial compression test. We simulate a laboratory-size specimen under plane strain, different confining pressures and with different residual friction angles. This example allows us to show the ability of the model to simulate the pressure dependence of the peak and residual strengths. We compare our numerical results with peak and residual strengths computed with a mechanical equilibrium model before and after the rupture.

\begin{figure*}
    \centering
	\makebox[2.0in][l] {\hspace*{0.1in}\textbf{(a)}~Biaxial model } \makebox[2.6in][l] {\hspace*{0.1in}\textbf{(b)}~Vertical force-displacement curves}\\
    \noindent\includegraphics[trim = 43mm 140mm 75mm 30mm, clip, width=2.0in]{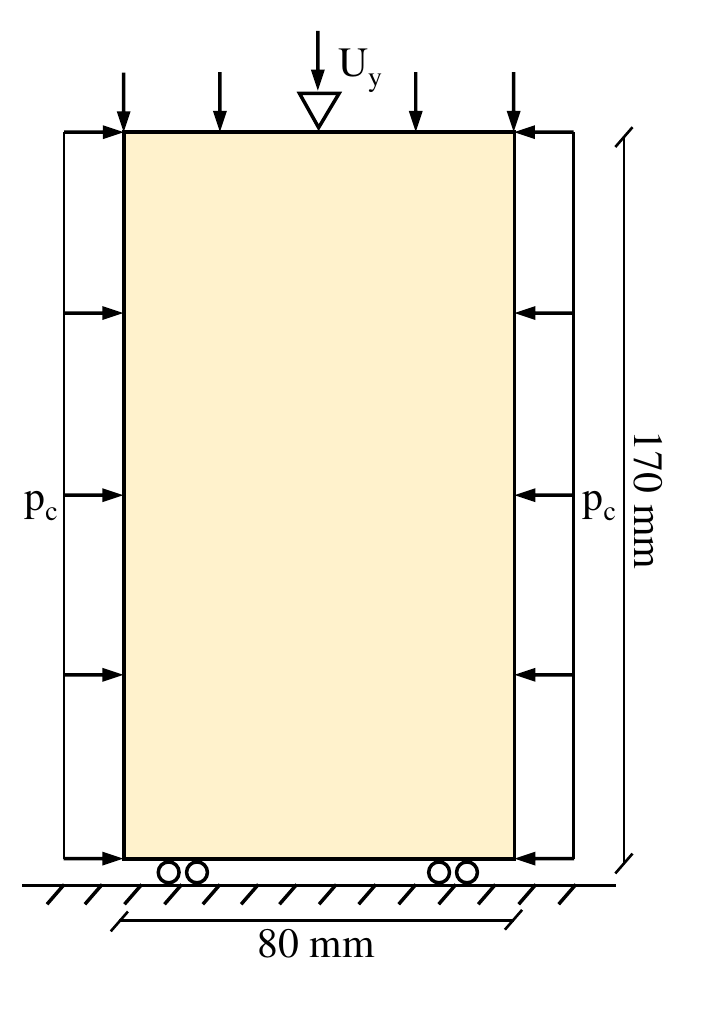} 
	\noindent\includegraphics[trim = 31mm 60mm 31mm 86mm, clip, width=2.6in]{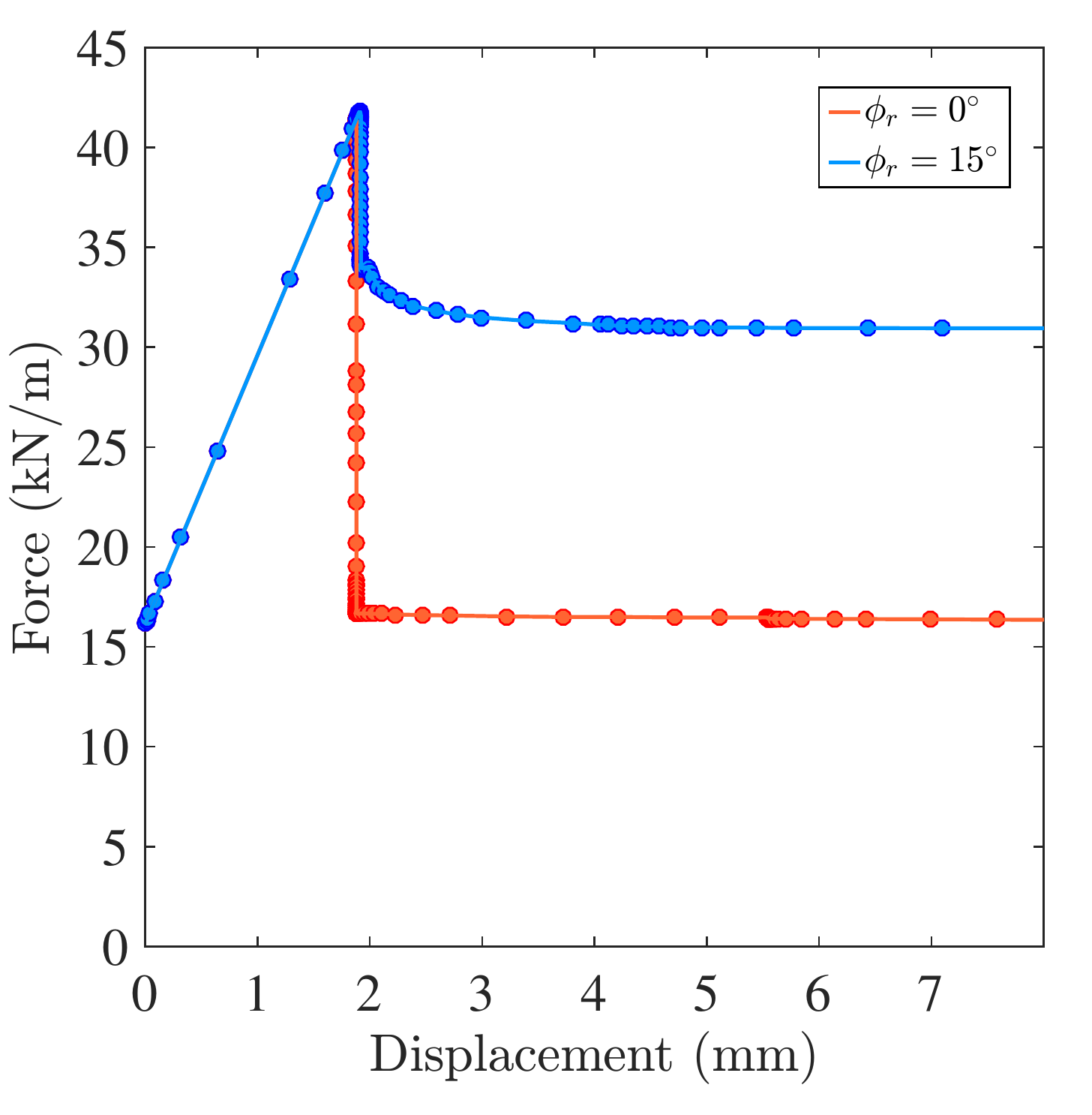} 
	\caption{Biaxial compression test. \textbf{(a)} The model setup. \textbf{(b)} Vertical force-displacement curves for two cases with $p_c=200$ kPa, $\phi=20^{\circ}$, and $\phi_r=20^{\circ}$ and $0^{\circ}$}\label{Fig1}
\end{figure*}

The model setup is shown in \Cref{Fig1}(a). The domain is 80-mm wide and 170-mm tall rectangular. The bottom boundary is supported by rollers, whereas a prescribed vertical displacement is imposed in the top boundary and zero horizontal displacement in the top middle point. The two lateral boundaries are subjected to the confining pressure, $p_c$, which is constant during the experiment. 

The material properties are: shear modulus $G=10$ MPa, Poisson's ratio $\nu=0.3$, cohesion strength $c=40$ kPa, peak friction angle $\phi=15^{\circ}$, shear fracture energy $\mathcal{G}_{c}=30$ J$/$m$^2$, and fracture's length-scale $l=2$ mm. We neglect gravity. We simulate three cases of $p_c$, 50 kPa, 100 kPa, and 200 kPa, and repeat each case with three values of the residual friction angles, $\phi_r=20^{\circ}$, $15^{\circ}$, and $0^{\circ}$. These simulations let us check whether our model captures the pressure dependence of the peak and residual strengths. We discretize the domain with a free triangular mesh with size $h=0.2$ mm that satisfy $l/h=10$.

We include two typical vertical force-displacement curves in \Cref{Fig1}(b). The confining pressure is $p_c=200$ kPa, the peak friction angle is $\phi=20^{\circ}$, and we consider two residual friction angles, $\phi_r=20^{\circ}$ and $0^{\circ}$. Initially, both vertical forces change linearly with the imposed vertical displacement until the peak strength is reached. The peak strength is the same in both models since they have the same $p_c$, $c$, and $\phi$. 
Afterward the fracture propagates suddenly across the domain, reaching both lateral boundaries, and the vertical force suddenly sinks. Our numerical model is able to capture the fracture propagation during the transition from the peak to the residual strengths due to the adaptive time step. Moreover, the curves evidence that the phase-field model is able to simulate the residual strength, which depends on the confining pressure, the residual friction angle, and the fracture path. 

\begin{figure*}
    \centering
	\noindent\includegraphics[trim = 32mm 65mm 37mm 85mm, clip, width=2.6in]{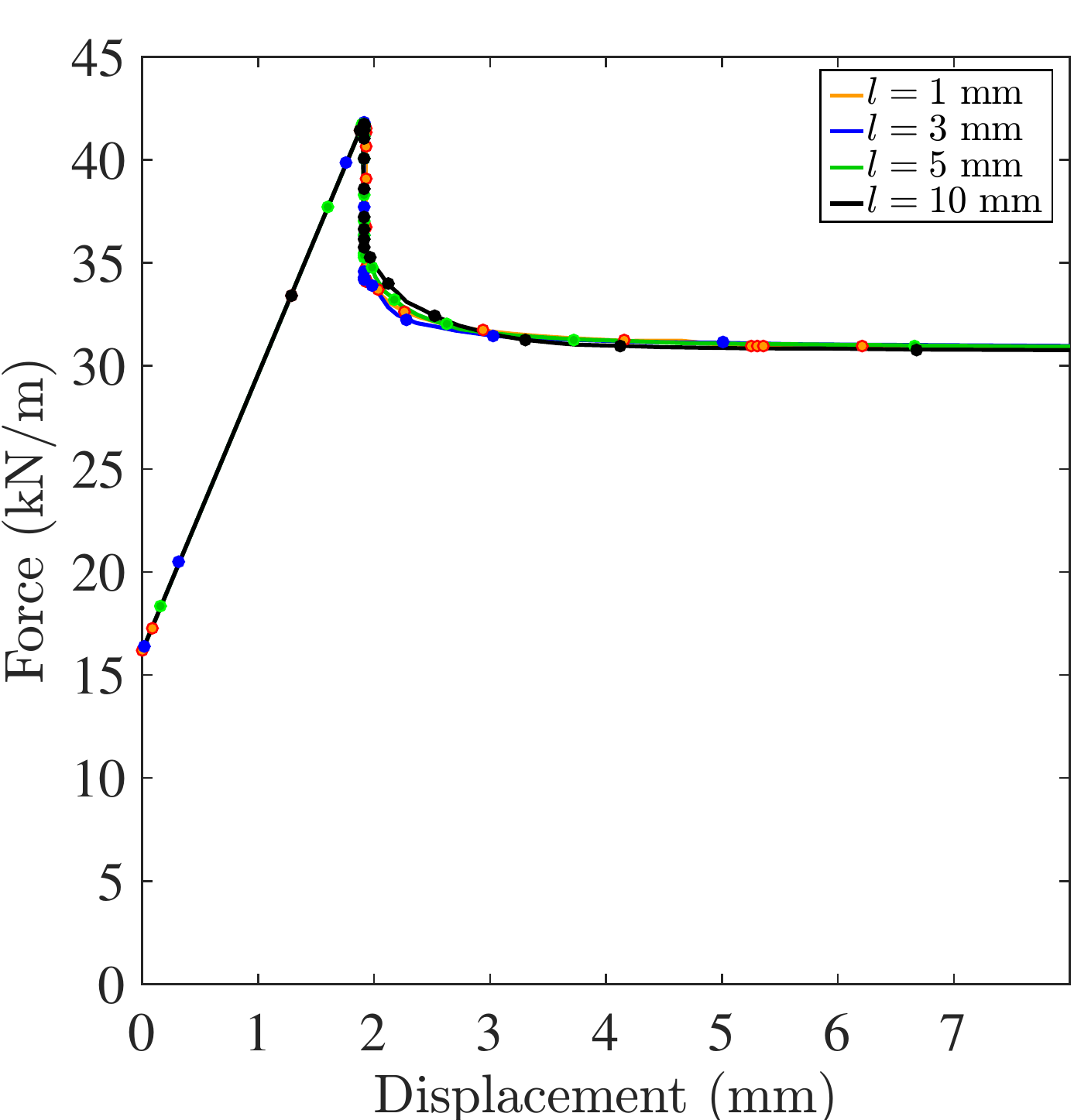} 
	\caption{Force-displacement curves with several phase-field length parameters, $l$.}\label{Fig1_rev}
\end{figure*}

We run several simulations of the biaxial compression problem for several values of $l$, ranging from 1 mm to 10 mm. The force-displacement curves for the four values of $l$ are included in Figure~\ref{Fig1_rev}. As in the previous problem, the curves for the values of $l$ confirm that the model is virtually insensitive to the phase-field length parameter.

\begin{figure*}
    \centering
	\makebox[1.5in][l] {\hspace*{0.1in}\textbf{(a)}~$U_y=1,9204$ mm } \makebox[1.5in][l] {\hspace*{0.1in}\textbf{(b)}~$U_y=1,9487$ mm} \makebox[1.5in][l] {\hspace*{0.1in}\textbf{(c)}~$U_y=2,9896$ mm}\\
    \noindent\includegraphics[trim = 105mm 5mm 105mm 5mm, clip, width=1.5in]{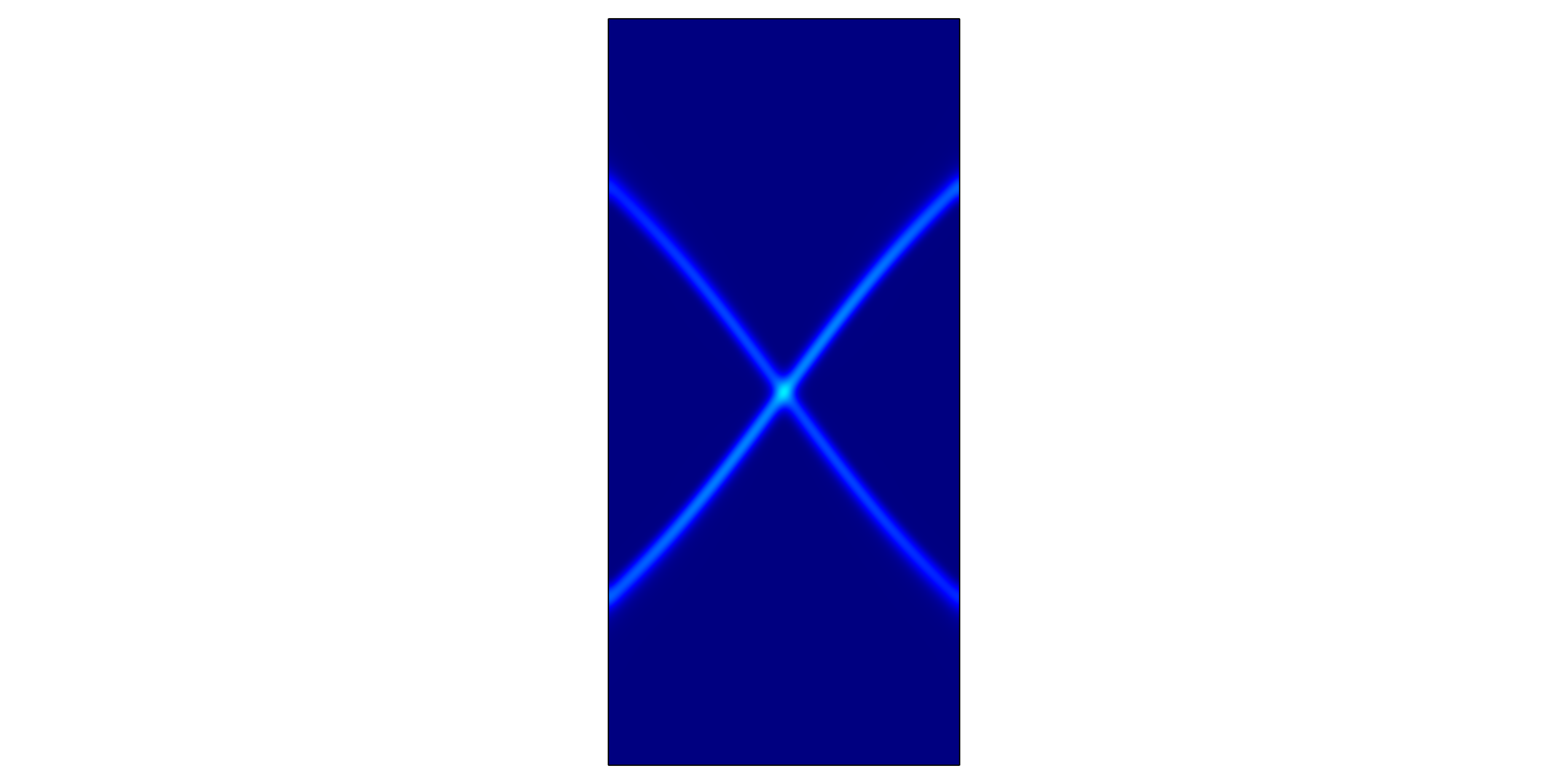} 
	\noindent\includegraphics[trim = 105mm 5mm 105mm 5mm, clip, width=1.5in]{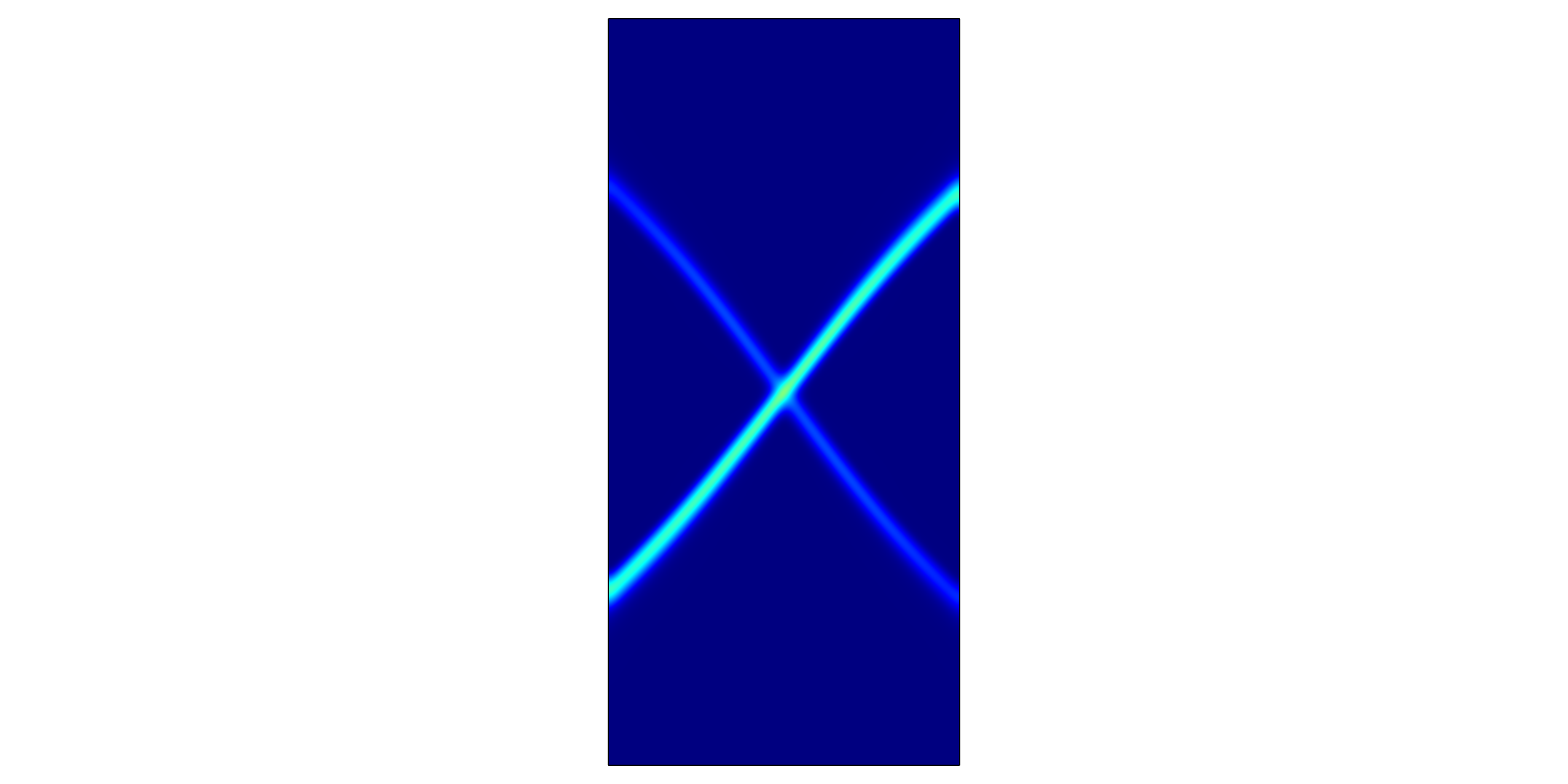} 
	\noindent\includegraphics[trim = 105mm 5mm 105mm 5mm, clip, width=1.5in]{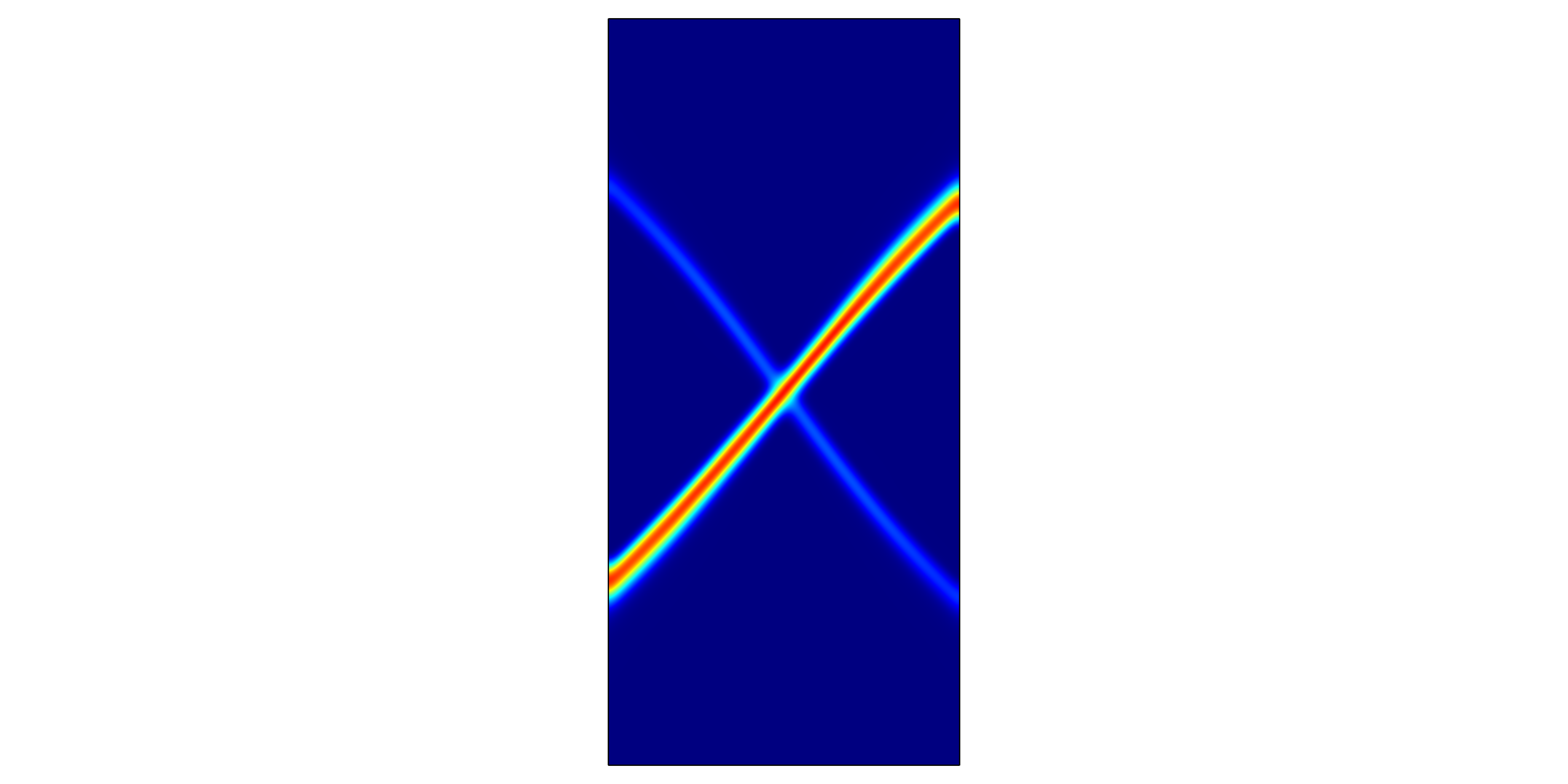} 
	\caption{Biaxial compression test. The evolution of the phase-field variable is plotted at three time steps. The confining pressure is $p_c=200$ kPa, the peak friction angle is $\phi=20^{\circ}$, and the residual friction angle $\phi_r=20^{\circ}$. The imposed vertical displacements, $U_y$, are: \textbf{(a)} $1,9204$ mm, \textbf{(b)} $1,9487$ mm, and \textbf{(c)} $2,9896$ mm.}\label{Fig2}
\end{figure*}

The evolution of the phase-field variable for $p_c=200$ kPa, $\phi=20^{\circ}$, and $\phi_r=20^{\circ}$, at three time steps is shown in \Cref{Fig2}. The phase-field is almost zero when the peak strength is reached, \Cref{Fig2}(a). In fact, due to the isotropic material model and homogeneous stress conditions of the biaxial test, two equally like fracture paths nucleate. This is consistent with the Mohr-Coulomb model. Nevertheless, only of the trajectories evolves and result in the final fracture pattern during the sudden decrease in the peak strength, \Cref{Fig2}(b). Later, the phase-field variable increases its value along the fracture path up to the residual peak strength is reached, \Cref{Fig2}(c).

\begin{figure*}
    \centering
        \noindent\includegraphics[trim = 66mm 93mm 72mm 67mm, clip, width=45 mm]{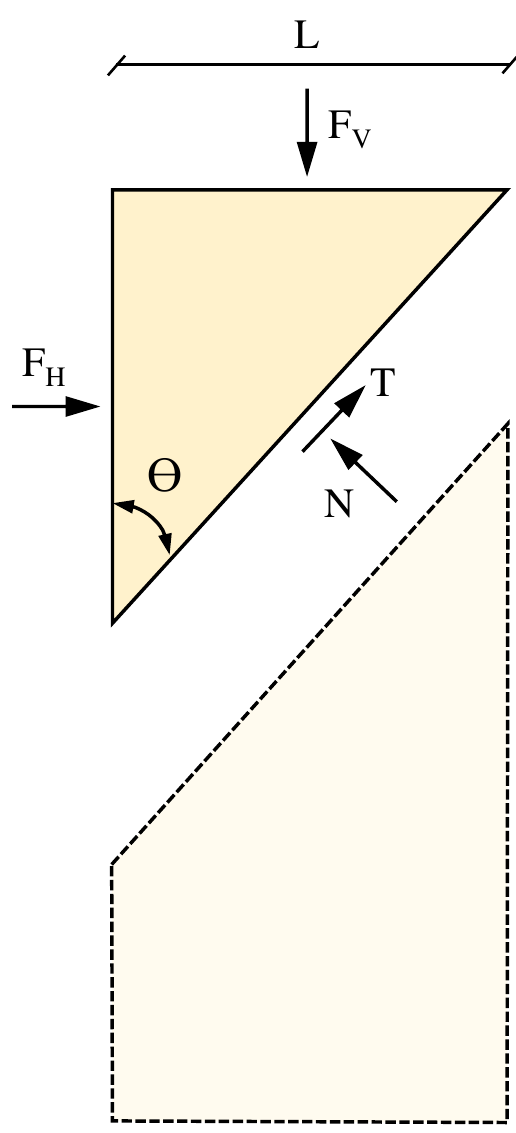} 
				\caption{Triaxial experiment. Mechanical equilibrium.}\label{Fig4}
\end{figure*}

We simulate nine cases with several combinations of $p_c$, $\phi$, and $\phi_r$ values. 
We also compute the peak and residual strengths applying mechanical equilibrium prior and after the fracture propagation. Given the fracture path, the mechanical equilibrium is illustrated in \Cref{Fig4}. The total vertical force applied on the top boundary is $F_V$, the total horizontal force on the left lateral boundary is $F_H$, and the tangential and normal forces on the fracture path are $T$ and $N$ respectively. We suppose the nucleation and fracture propagation is instantaneous and the fracture path is a straight line. the angle between the fracture path and the vertical axis is~$\theta$. Then, at the onset of the fracture propagation, the tangential force on the fracture is:
\begin{equation}\label{eq:Tb}
    T=\frac{L}{\sin \theta}c+N\tan \phi,
\end{equation}
and once the fracture is fully developed, the tangential force on the fracture is:
\begin{equation}\label{eq:Te}
    T=N\tan \phi_r.
\end{equation}

The mechanical equilibrium in the vertical direction is given by:
\begin{equation}\label{eq:Fv}
    V-T \cos \theta -N \sin \theta=0,
\end{equation}
and in the horizontal direction:
\begin{equation}\label{eq:Fh}
    H+T \sin \theta -N \cos \theta=0,
\end{equation}
where $H$ is:
\begin{equation}\label{eq:H}
    H=p_c \frac{L}{\tan \theta}.
\end{equation}

Solving $V$ from Eq.~\eqref{eq:Fv}, substituting $V$ in Eq.~\eqref{eq:Fh} and operating, the vertical force at the onset of the fracture propagation $V_p$ --peak strength-- is:
\begin{equation}\label{eq:Vpeak}
    V_p=\frac{1}{\cos \theta -\sin \theta \tan \phi} \left(  \frac{L \cdot c}{\sin \theta} +p_c \frac{ L}{\tan \theta} \left( \cos \theta tan \phi + \sin \theta \right) \right),
\end{equation}
and the vertical force once the fracture is fully propagated $V_r$ --residual strength-- is:
\begin{equation}\label{eq:Vresi}
    V_r=\frac{1}{\cos \theta -\sin \theta \tan \phi_r} \left( p_c \frac{ L}{\tan \theta} \left( \cos \theta tan \phi_r + \sin \theta \right) \right).
\end{equation}

We compute $V_p$ and $V_r$ for the nine simulated cases. The results are listed in Table~\ref{tab:resTen}. The agreement between both models is remarkable.

\begin{landscape}
\begin{table} 
\centering
\begin{tabular}{lcccccccccccc} \hline &\\ [-.4cm]

& \multicolumn{4}{c}{$\phi=15^{\circ}$ $\phi_r=15^{\circ}$} &	\multicolumn{4}{c}{$\phi=20^{\circ}$ $\phi_r=20^{\circ}$} &	\multicolumn{4}{c}{$\phi=15^{\circ}$ $\phi_r=0^{\circ}$}	\tabularnewline &\\ [-.4cm]

& \multicolumn{2}{c}{Peak stre.}& \multicolumn{2}{c}{Residual stren.} & \multicolumn{2}{c}{Peak stren.} & \multicolumn{2}{c}{Residual stren.} & \multicolumn{2}{c}{Peak stren.}& \multicolumn{2}{c}{Residual stren.} \tabularnewline &\\ [-.4cm]

& M. Eq. & Sim. & M. Eq. & Sim. & M. Eq. & Sim. & M. Eq. & Sim. & M. Eq. & Sim. & M. Eq. & Sim. \tabularnewline &\\ [-.4cm]	

$p_c=50$ kPa & $15.19$ &	$15.15$ &	$6.81$ &	$7.12$ &	$17.30$ &	$17.30$ &	$8.16$ &	$8.24$ &	$15.67$ &	$15.12$ &	$4.00$ &	$4.17$ \tabularnewline &\\ [-.4cm]	

$p_c=100$ kPa & $22.00$ &	$21.94$ &		$13.62$ &		$14.27$ &		$25.46$ &		$25.56$ &		$16.32$ &		$17.04$ &		$22.59$ &		$21.93$ &		$8.00$ &		$8.23$ \tabularnewline &\\ [-.4cm]	

$p_c=200$ kPa & $35.62$ &	$34.67$	 & $27.23$ &	$27.80$ &	$41.77$ &	$41.48$ &	$32.63$ &	$33.17$ &	$36.45$ &	$35.48$ &	$16.00$ &	$16.30$ \tabularnewline \hline &\\ [-.4cm]	
	
\end{tabular}
\caption{Triaxial experiment. We list the peak and residual strengths for nine cases in kN. Both strengths are computed with our phase-field model --denoted as Sim.-- and using a mechanical equilibrium --denoted as M. Eq.-- prior and after the fracture propagation.}
\label{tab:resTen}
\end{table}
\end{landscape}

\subsection{Slope failure analysis}

\begin{figure}
    \centering
    \includegraphics[width=1.\textwidth]{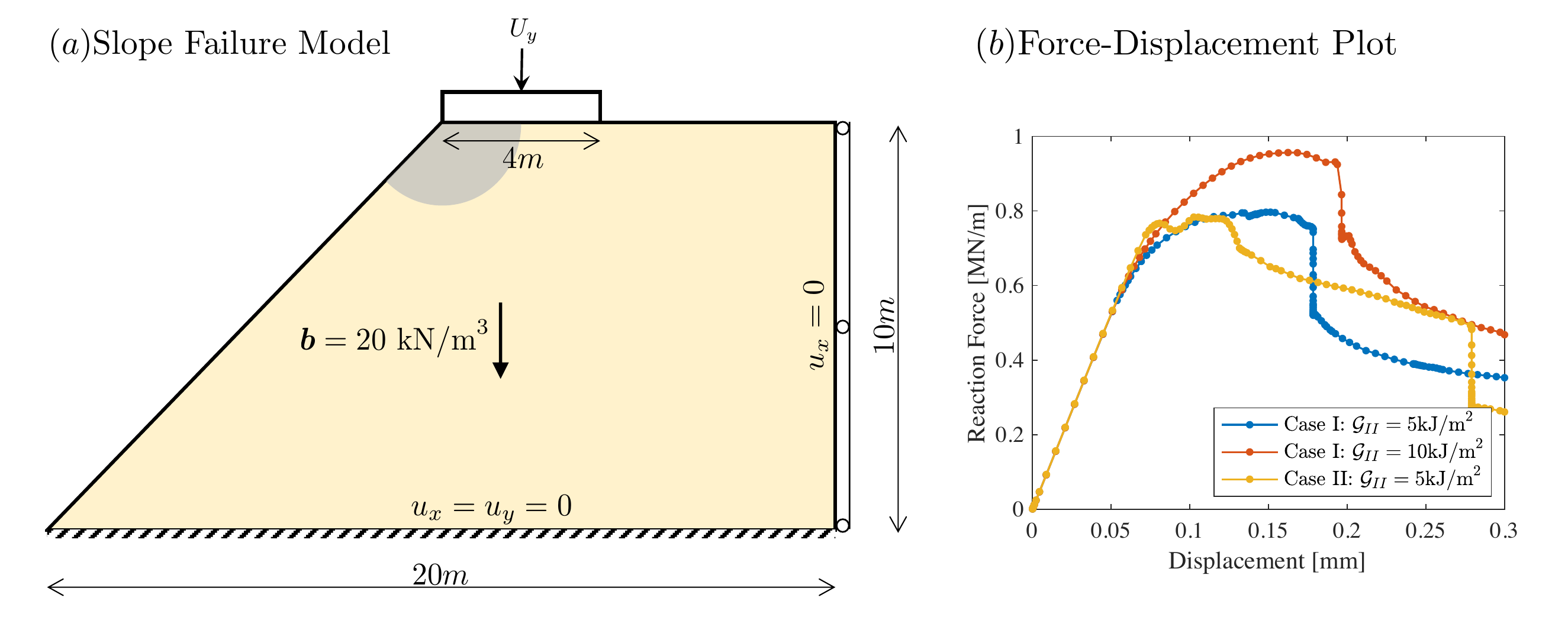}
    \caption{Slope failure analysis. (a) The model setup for the slope failure analysis. The domain is 20 m wide and 10 m tall, with a slope 1:1. A 4 m wide rigid footing is located on the slope's crest, and is subjected to a vertical displacement $U_y$. The boundary conditions are: bottom edge is fixed in both directions, right edge is fixed horizontally and other faces are traction free. The grey region highlights the damage-inactive (Case I) and damage-active (case II) problems. (b) Force displacement results measured at the point of loading in the middle of the footing. The plots show absolute values. }
    \label{fig:fig8}
\end{figure}

As the last example, we consider the problem of slope failure analysis reported in \cite{regueiro2001plane}. Consider the soil slope shown in \Cref{fig:fig8}. The domain is 20 m wide and 10 m tall, with a slope 1:1 on the left side. A 4 m wide rigid footing is placed on the crest of the slope. 
The slope is first subjected to a body force $b=20~\text{kN/m}^3$, and then these body-force stresses are used as the initial state for the footing loading step. 
Displacement at the bottom edge is fixed in both directions, while for the right edge, only horizontal displacement is fixed. 
As the main loading step, a displacement $U_y=0.3~\text{m}$ is prescribed in the middle of a rigid foundation, which simulates the effect of a building imposing a stress on the slope. 

The elastic parameters of the soil include $E=10~\text{MPa}$ and $\nu = 0.4$. The initial friction angle and cohesion are $\phi = 16.7^\circ$ and $c=40~\text{kPa}$, with $\phi_r=10^\circ$ and $c_r = 0~\text{kPa}$ as their respective residual values. The phase-field length scale parameter is set to $l = 200~\text{mm}$, and the domain is discretized using a free triangular mesh with mesh-size $20~\text{mm}$. The resulting mesh roughly has 1M triangles and 500K vertices. The computational time takes about 12 hours in our desktop machine with i9-10900 processor with 10 cores and 20 threads.

\begin{figure}
    \centering
\includegraphics[width=1\textwidth]{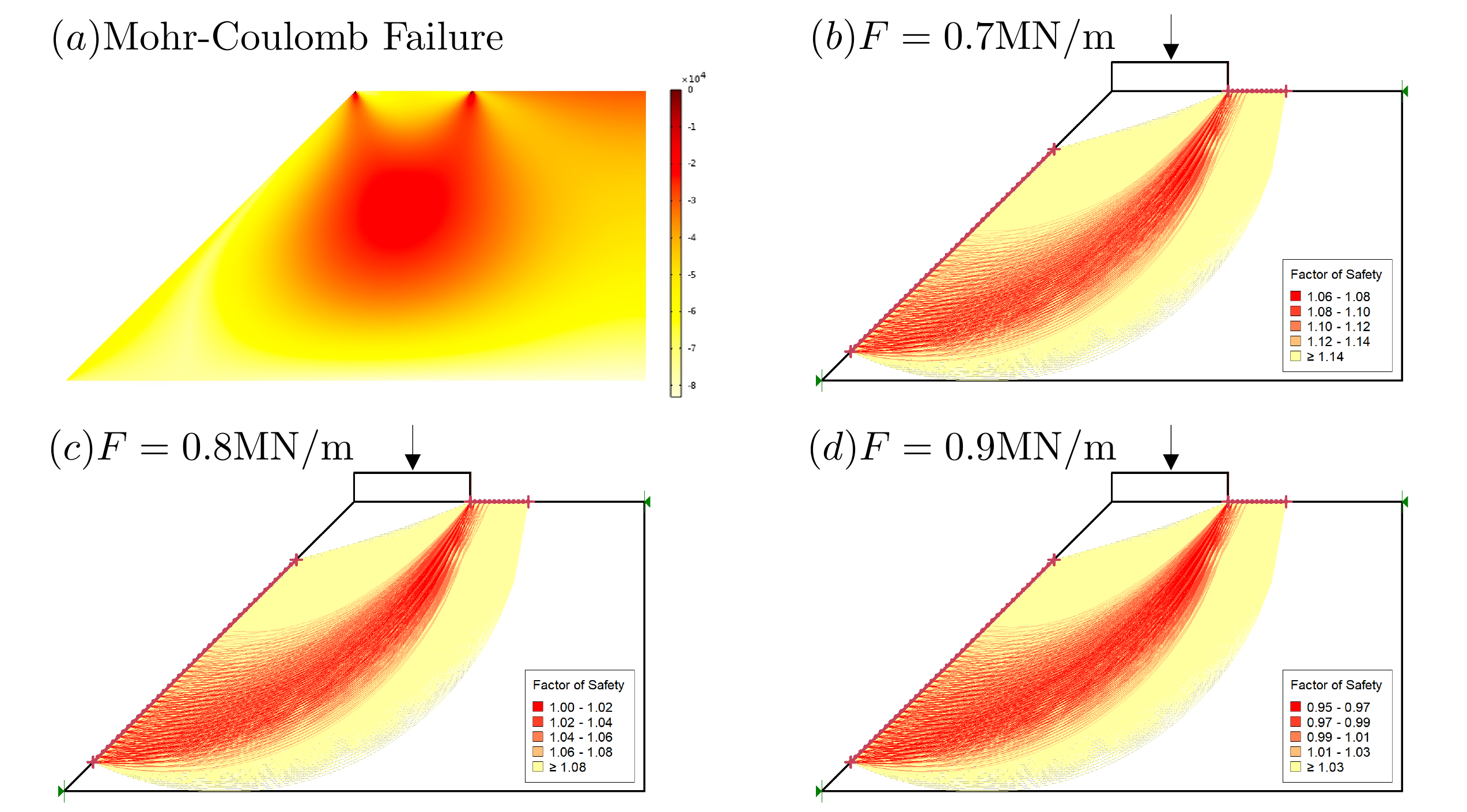}
\caption{(a) Mohr-Coulomb failure function plotted right before the onset of localization. The stress around both corners of the footing appear as near-failure critical. (b-d) Mohr-Coulomb's critical surfaces using limit-equilibrium method at different load. $F\approx0.8\text{MN/m}$ is found as the critical load, where factor of safety of the slope reaches 1. }
    \label{fig:fig9}
\end{figure}

\begin{figure}[H]
    \centering
    \includegraphics[width=1.\textwidth]{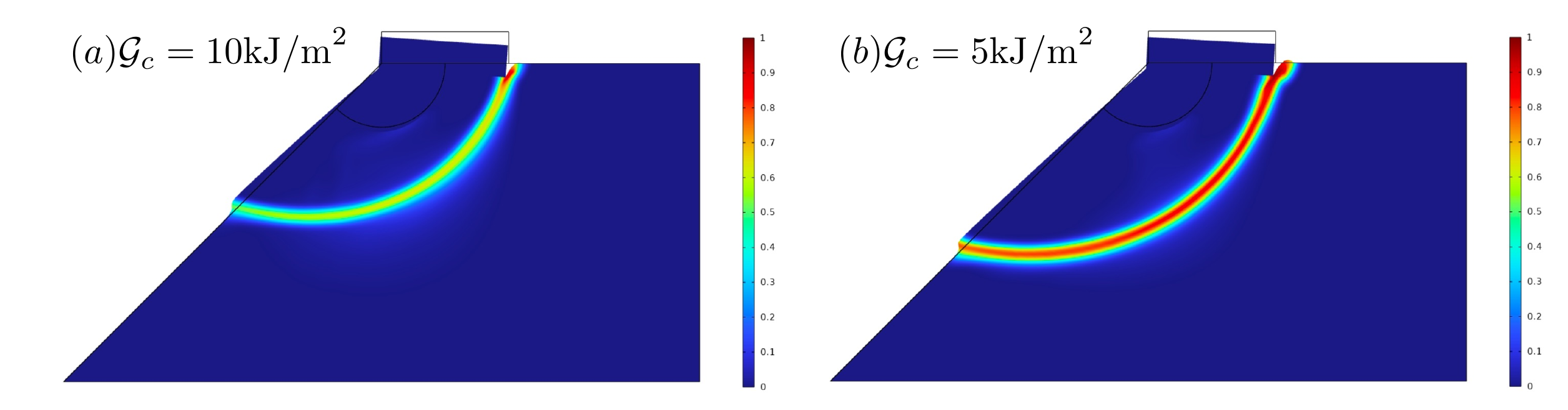}
    \caption{Damage evolution for the case I slope stability analysis, where the left side is suppressed to have damage development. The plots show the evolution damage at $U_y=300\text{mm}$ for (a) $\mc{G}_c=10\text{kJ/m}^2$ and (b) $\mc{G}_c=5\text{kJ/m}^2$. }
    \label{fig:fig10}
\end{figure}

Due to the relatively high cohesion and low friction angle, the shear-band formation for this problem is particularly interesting. If we plot the evolution of the Mohr-Coulomb’s failure envelope right before the onset of fractures, as shown in \Cref{fig:fig9}-(a), we observe that the failure should onset from both ends of the footing. This fact has also been reported by \citet{haghighat2015modeling}, however, due to the pre-specification of only one orientation angle ($\theta$), the crack formation from the left side was not captured by \citet{fei2020phase}. 
Therefore, to perform a comparison, we consider two cases:
\begin{itemize}
    \item[I.] Shear band formation only from the right corner of the footing by suppressing the phase field variable to zero ($d=0$) in the gray region (see \Cref{fig:fig8}).
    \item[II.] Free shear-band formation, which results in two patterns from each side followed by coalesces. 
Additionally, we consider two critical fracture energies of $\mc{G}_c = 10 \text{kJ/m}^2$ and $\mc{G}_c = 5 \text{kJ/m}^2$. The final fracture patterns of these two cases are shown in \Cref{fig:fig10}. 
\end{itemize}

\begin{figure}[H]
    \centering
    \includegraphics[width=1.\textwidth]{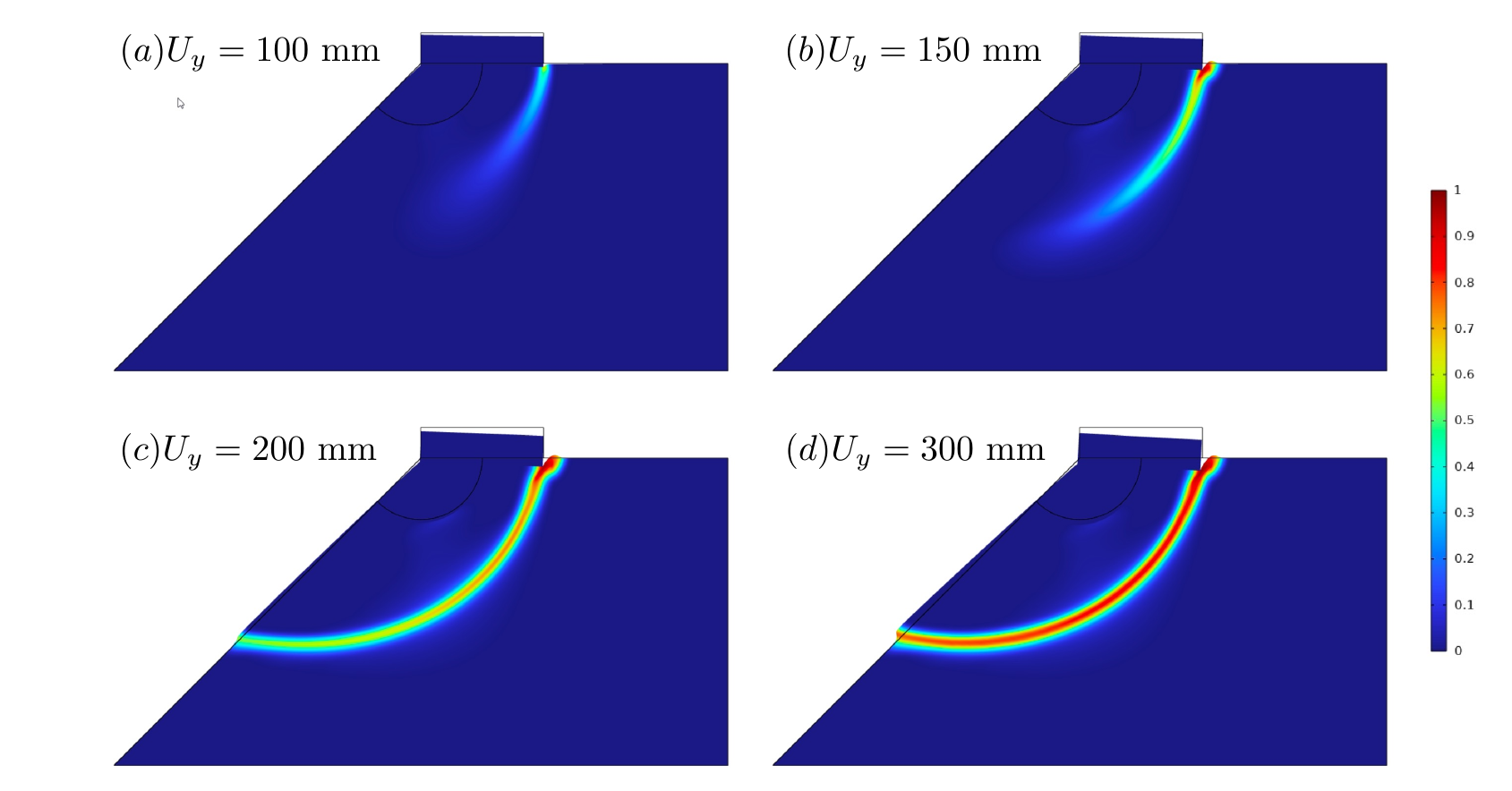}
    \caption{Damage evolution for the case I slope stability analysis ($\mc{G}_c=5\text{kJ/m}^2$). In this case, the left side is suppressed to develop damage, therefore we have a single crack formation from the footing's right corner. Subplots a-d show the evolution of the damage parameter at different loading steps. }
    \label{fig:fig11}
\end{figure}

The evolution of phase-field variable for case I, with $\mc{G}_c = 5 \text{kJ/m}^2$, are plotted for different loading steps in \Cref{fig:fig11}. The force-displacement response is plotted in \Cref{fig:fig8}-(b). 
As the reader can find, the proposed formulation captures the peak and residual loads as well as the crack patterns accurately, and the results are consistent with those reported by \citeauthor{fei2020phase}. The failure surface evaluated using phase-field method and the peak-load is well-aligned with potential failure surfaces and critical load $F=0.8\text{MN/m}$ resulting from  limit-equilibrium analysis of the slope using the GeoStudio software (see \Cref{fig:fig9}-b-d).

\begin{figure}[H]
    \centering
    \includegraphics[width=1.\textwidth]{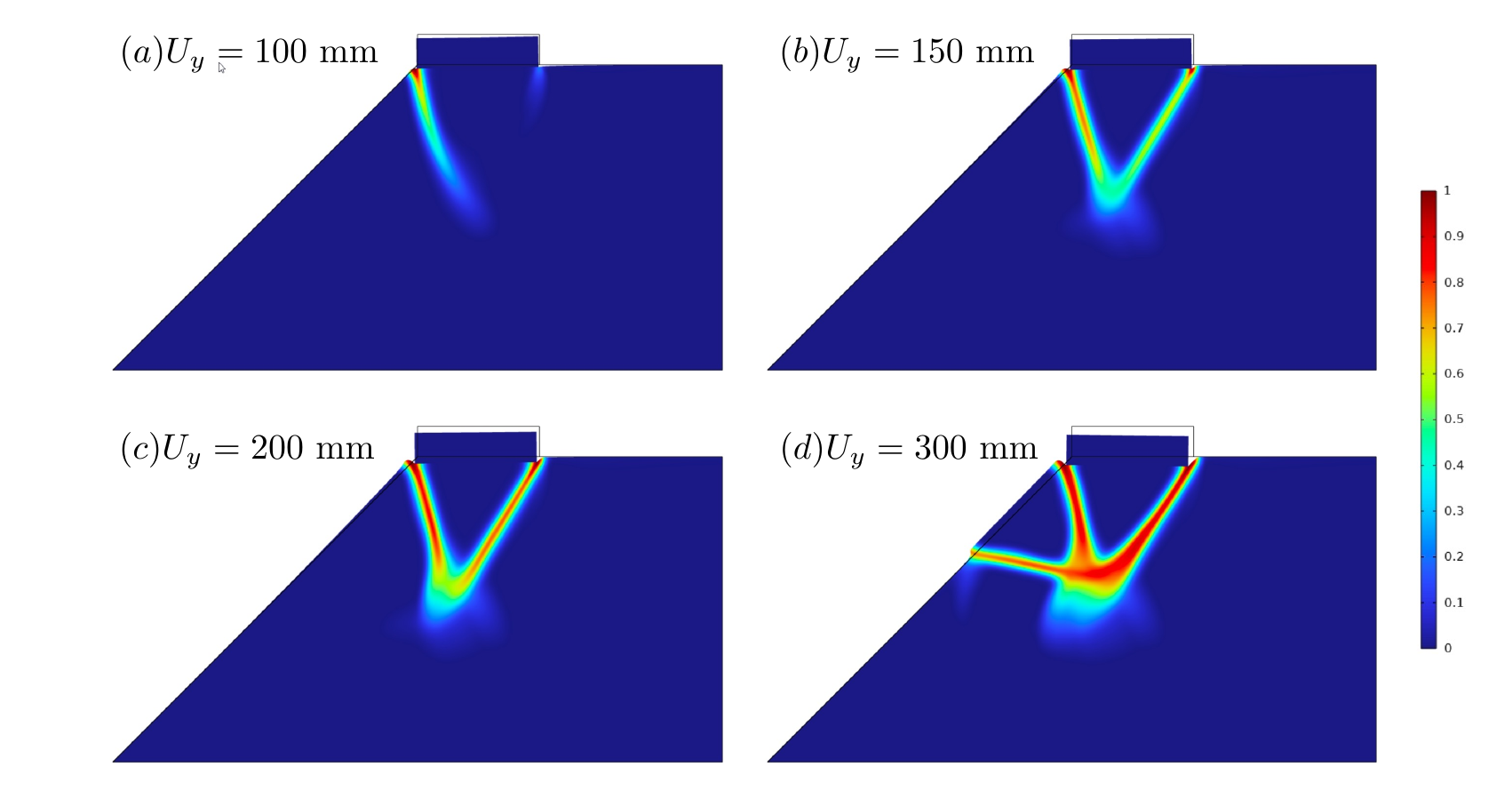}
    \caption{Damage evolution for the case II slope stability analysis ($\mc{G}_c=5\text{kJ/m}^2$). In this case, the damage can initiate from either side and is primarily driven by the crack energy. Subplots a-d show the evolution of the damage parameter at different loading steps.}
    \label{fig:fig12}
\end{figure}

Lastly, we run a new set of simulations for case II. The results are plotted in \Cref{fig:fig12}. As we find, here the model captures first a shear band formation from the left corner of the footing. This is in fact expected because of the stress-free surface of the slope creates a more critical failure condition on the left corner. The propagation of the mode, however, stops because it is directing to Mohr-Coulomb stable regions of the domain. Later, the main failure mode initiates and propagates from the right corner, and collides with the first mode somewhere underneath the footing, which is also consistent with the results of the limit state theories. A final branch is then generated and causes the ultimate failure of the slope. The pick stress, however, does not seem to be very different from those of Case I, as plotted in \Cref{fig:fig8}-(b).

\section{Concluding remarks}

We presented a phase-field model of shear fractures using deviatoric stress decomposition (DSD). We validated the model by solving reference problems of shear fractures in geotechnical engineering. Our model has excellent performance. 
The main advantages of our phase-field approach are: (1) the model does not require re-meshing, (2) nucleation, propagation, and fracture path are automatically computed without the need to track fractures or pre-specify orientations, and (3) fracture joining and branching do not need additional algorithms.

For an isotropic Mohr-Coulomb material under homogeneous loading, it has been shown that there are two conjugate surfaces having the same likelihood for shear band formation. In fact, our model captures this for the biaxial compression problem without any intervention.
This is the same for the slope stability problem, where our model was able to capture crack initiation from both corners of the foundation.
While accurate in peak and residual force calculations, 
we found that the CSD model of shear fractures is more accurate in capturing such a transition. 

The study was limited to modeling two-dimensional problems of compressive fracture. However, the proposed formulation is not limited to any dimensions. Therefore, we plan to explore three-dimensional models as a follow-up study. Additionally, pore-fluid consideration is critically important for modeling failure in geomaterials. This is also an area that will be considered next. Additional paths include incorporating rate-and-state friction models that are best suited for modeling geologic systems, and thermal coupling that is important for modeling geothermal systems.

\appendix
\section{Crack driving force for deviatoric stress decomposition}\label{appen1}

Reminding that $\hat{q}, \tilde{q}_p, \tilde{q}_r$ denote the bulk and fractured deviatoric stresses at peak and residual stages, respectively, and $\hat{q} = 3\mu\varepsilon_q$, with $\varepsilon_q$ and the deviatoric strain, the crack driving force during a plastic dissipation process as a result of frictional sliding can be expressed as 
\begin{equation}\label{eqs:A1}
\begin{split}
\mc{H}_{slip} & = \int_{\varepsilon_{\gamma}^p}^{\varepsilon_{\gamma}} (\hat{q} - \tilde{q}_r) d\varepsilon_{\gamma} \\
& = \int_{\varepsilon_{\gamma}^p}^{\varepsilon_{\gamma}} (3\mu\varepsilon_q - \tilde{q}_r) d\varepsilon_{\gamma} \\
& = \frac{1}{6\mu}\left(\hat{q}^2 - \tilde{q}_p^2\right) - \frac{\tilde{q}_r}{3\mu}\left(\hat{q}- \tilde{q}_p \right) \\
& = \frac{1}{6\mu}\left\{(\hat{q} - \tilde{q}_r)^2 - (\tilde{q}_p - \tilde{q}_r)^2 \right\}.
\end{split}
\end{equation}
Since $q = g(d)\hat{q} + (1-g(d))\tilde{q}_r$, we will have,
\begin{equation}\label{eqs:A2}
\begin{split}
\mc{H}_{slip} & = \frac{1}{6\mu} \left\{ \left(\frac{q - \tilde{q}_r}{g(d)}\right)^2 - (\tilde{q}_p - \tilde{q}_r)^2 \right\}.
\end{split}
\end{equation}
We observe that the relations are quite similar to those reported by \citeauthor{fei2020phase} using shear stress split, except that shear stresses and strains are replaced now with deviatoric ones and therefore division by $3\mu$ instead of $\mu$. 

Noting that total driving energy is expressed as $\mc{H} = \mc{H}_t + \mc{H}_{slip}$, re-arranging \cref{eqs:A2}, we can write
\begin{equation}\label{eqs:A3}
\begin{split}
\mc{H} = \left\{\mc{H}_t - \frac{1}{6\mu}(\tilde{q}_p - \tilde{q}_r)^2\right\} + \frac{1}{6\mu} \left(\frac{q - \tilde{q}_r}{g(d)}\right)^2.
\end{split}
\end{equation}
Now, one can substitute this relation into the phase field PDE \cref{pde_d}, and with 1D simplifications, integrate the phase field relation, as detailed in \cite{fei2020phase}, to arrive at approximate relations for the evolution of deviatoric stress $q$ as a function damage. Again, since the phase field PDE \cref{pde_d} and driving force \cref{eqs:A3} are very similar to those in \cite{fei2020phase}, all the derivations hold identical and true for the deviatoric stress decomposition. Finally, by imposing length-scale independency to the deviatoric stress evolution, one obtains that 
\begin{equation}\label{eqs:A4}
\begin{split}
\mc{H}_t = \frac{1}{6\mu}(\tilde{q}_p - \tilde{q}_r)^2.
\end{split}
\end{equation}
This completes the derivation of crack driving force relations introduced in \cref{eqs:28,eqs:29}.

\section*{Data Availability Statement}

All data, models, or code generated or used during the study will be made available online at \href{https://github.com/ehsanhaghighat/PhaseField-DSD}{https://github.com/ehsanhaghighat/PhaseField-DSD} upon publication.

\section*{Acknowledgements}
This research Project has been funded by the Comunidad de Madrid through the call Research Grants for Young Investigators from Universidad Politécnica de Madrid under grant APOYO-JOVENES-21-6YB2DD-127-N6ZTY3, RSIEIH project, research program V PRICIT. Authors acknowledge the help of Mrs. Aida Rezapour (M.Sc., P.Eng.) in preparing slope stability results using the limit equilibrium method.

\end{document}